\documentclass[reprint, amsmath, amssymb, aps]{revtex4-1}

\pdfoutput=1

\usepackage{graphicx}
\graphicspath{{figs/}}
\usepackage{dcolumn}
\usepackage{bm}
\usepackage{amsmath}
\usepackage{subcaption}
\usepackage{color}

\begin{document}

\preprint{APS/123-QED}

\title{The supremum principle selects simple, transferable models}

\author{Cody Petrie}
 \email{codypetrie89@gmail.com}
\author{Christian Anderson}
\author{Casie Maekawa}
\author{Travis Maekawa}
\author{Mark K. Transtrum}
 \email{mktranstrum@byu.edu}
 \affiliation{
 Department of Physics and Astronomy\\Brigham Young University\\Provo, UT 84602
}

\date{\today}

\begin{abstract}
    We consider how mathematical models enable predictions for conditions that are qualitatively different from the training data.
    We propose techniques based on information topology to find models that can apply their learning in regimes for which there is no data.
    The first step is to use the Manifold Boundary Approximation Method to construct simple, reduced models of target phenomena in a data-driven way.
    We consider the set of all such reduced models and use the topological relationships among them to reason about model selection for new, unobserved phenomena.
    Given minimal models for several target behaviors, we introduce the \emph{supremum principle} as a criterion for selecting a new, transferable model.
    The supremal model, i.e., the least upper bound, is the simplest model that reduces to each of the target behaviors.
    We illustrate how to discover supremal models with several examples; in each case, the supremal model unifies causal mechanisms to transfer successfully to new target domains.
    We use these examples to motivate a general algorithm that has formal connections to theories of analogical reasoning in cognitive psychology.
\end{abstract}

\maketitle

One of the first important tasks in modeling data is selecting the form for a mathematical model.
The form of the model defines the types of predictions a model can make and therefore--accurately or not--creates a type of ``hypothesis space'' called \emph{inductive bias} \cite{baxter2000model}.
In this study, we use the geometric and topological relationships among candidate models to reason about inductive bias and model selection.
Of particular interest are predictions for qualitatively different conditions than those on which a model was trained, such as predicting a time series outside of the range of sampled time points, predicting under different experimental conditions, or applying insights from two populations to a third.
A model's ability to make such \emph{out-of-domain} predictions is sometimes known as \emph{transferability}, which is stronger than generalization, i.e., predicting data generated for inputs similar to those on which it was trained \cite{weiss2016survey}.
We propose a general principle of model selection, the \emph{supremum principle}, that encodes a preference for simplicity with respect to target quantities of interest while enabling model transferability and whose construction uses topological relationships formally equivalent to models of human analogical reasoning.

One of the key struggles of model selection is balancing inductive bias against model flexibility.
Consider, for example, explaining the change in a cell's state (e.g., healthy to cancerous) in terms of the proteome.
A potential hypothesis space could include all possible interactions between all $\sim$25,000 known proteins.
This has very little bias since the correct explanation is somewhere in this space; however, it would require an unreasonable amount of data to learn all the parameters of such a complex model.
Furthermore, it would be even more difficult to interpret the model afterwards as most of the interactions are merely explanatory noise relative to the phenomenon of interest \cite{batterman2001devil}.
Therefore, we seek to restrict the hypothesis space to the one that minimally includes our behavioral regimes of interest.
Such a model doesn't fit both states with one set of parameters, rather, it fits either set of data independently, i.e. some parameters could be unidentifiable to data from either state.
In addition to describing cells in either state, this model \emph{predicts} a mechanism for switching between them.

A minimal criterion for a useful predictive model is that it reproduces the training data within statistical noise, that is, a kind of coarse interpolation.
Common statistical practices such as holdout, jackknife, and cross-validation reinforce this intuition.
Sloppy models \cite{brown2003statismechan, brown2004statismechan, waterfall2006sloppmodel, machta2013paramspace}, a class of over-parameterized models, further formalize the relation between prediction and interpolation using information geometry \cite{amari2007methods, transtrum2010}.
The predictions of sloppy models are controlled by only a few \emph{stiff} parameter combinations and so are said to have a \emph{low effective dimensionality}\cite{transtrum2010, lamont2019correspondence}.
Effective dimensionality is quantified in terms of widths of a model manifold,  rigorous bounds for which are given by theorems from interpolation theory \cite{transtrum2010, quinn2019}.
Indeed, it has been suggested that predictive models are generalized interpolation schemes \cite{transtrum2011geometnonlin}.

\begin{figure*}
\includegraphics[width=\textwidth]{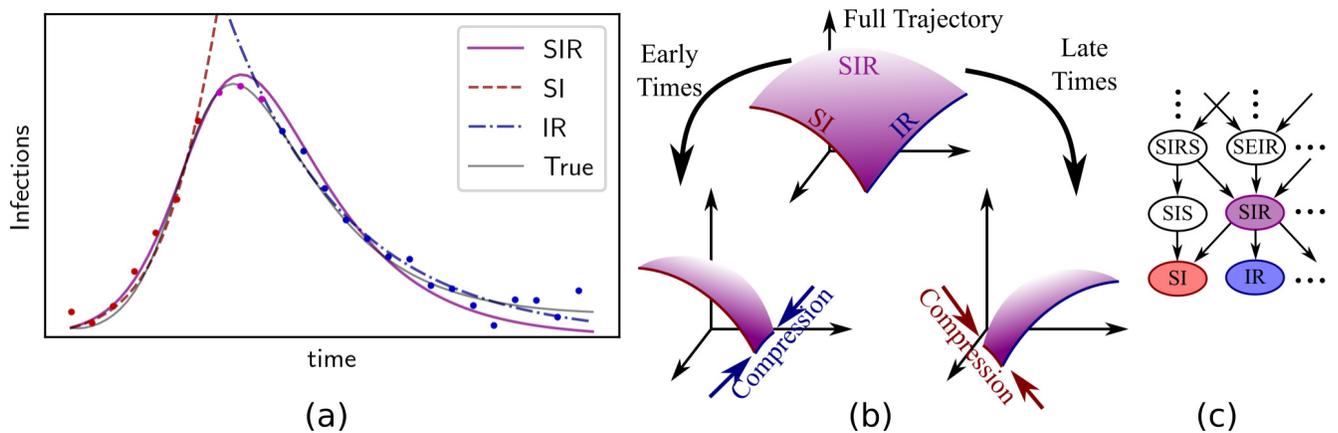}
\caption{(a) Infections versus time for a hypothetical epidemic.
  Data taken at different times---early (red), late (blue), and intermediate (purple)---exhibit qualitatively different types of behaviors.
  These data carry information about different aspects of the generating process.
  The ``True" curve represents the model from which data was generated, without the corruption of Gaussian noise.
  (b) Information geometry quantifies how data identify a model's parameters.
  The full trajectory completely identifies an SIR model, manifest by a model manifold that is not compressed along either direction.
  Restricting to data at early times compresses the directions related to a recovery rate so that the model is well-approximated by an effective SI model.
  Similarly, restricting to late times compresses information about infection, leading to an effective IR model.
  (c) Candidate models of varying complexity can be arranged hierarchically in a directed graph.
  The SIR model is the \emph{supremum} of the SI and IR models, i.e., the simplest model that combines information from both early and late times.
  It can use the information from data in two of the regimes to accurately predict a third.
}\label{SIR}
\end{figure*}

However, there is a sense that more than simple interpolation ought to be possible \cite{lake2017building, webb2020learning}. 
Human cognition is driven by understanding, rather than mere pattern mimicry.
When we reason about molecular bonds as if they were balls and springs, we use analogical reasoning to identify abstract relationships and transfer insights among superficially different systems.
Can machines similarly analogize to make predictions of a qualitatively different nature than those on which they were trained?

To explore this question, we use information geometry to assess parameter identifiability and predictive performance for models fit to data from different regimes and reason about the hypotheses they encode.
The Fisher Information Matrix (FIM) is information geometry's fundamental object, a Riemannian metric on a manifold of models using parameters as coordinates \cite{transtrum2010, brouwer2018underlying}.
Model manifolds are often thin, and boundaries correspond to simplified models, i.e., having fewer parameters \cite{transtrum2014modelreduc}.
Distances measured by the FIM typically \emph{compress} the model manifold into a few relevant directions \cite{machta2013paramspace} so that the manifold is thin and well-approximated by a low-dimensional, simplified model that resides on the boundary.
Given training data, the Manifold Boundary Approximation Method (MBAM) explicitly finds limiting approximations to give a minimal, reduced model that encodes the information in the data.
MBAM is an enabling technology for our approach and is described in detail elsewhere references\cite{transtrum2014modelreduc,transtrum2015perspective}.

Given several reduced models for target quantities of interest, we next seek a single model that unifies their simplified explanations. 
To choose an appropriate model, we introduce the \emph{supremum principle}: select the simplest model that is reducible to each of the target behaviors.
One of the primary contributions of this paper is to show that this intuitive idea can be given a rigorous definition using the formalism of information topology.
We call this model the \emph{supremal model} and give an algorithm below for constructing it.
The supremum principle formally encapsulates a preference for simplicity akin to Occam's razor, motivated by the assumption that abstract models that explain multiple behaviors are more likely to transfer accurately to novel behaviors than models developed for a single phenomenon.

As a motivating example, consider modeling infection trajectories during an epidemic.
Fig.~\ref{SIR}a shows data generated from an MSEIR model with birth and death rates (six parameters, fifth-order dynamics) and corrupted by Gaussian noise.
We partition the data into three qualitatively distinct regimes---early (red), intermediate (purple), and late (blue)---and ask: Which subsets of the data are informative for predicting data in another regime?

To illustrate the key principles, consider fitting the data with a simple SIR model (two parameters, second-order dynamics),
\begin{align}
  \label{eq:SIR}
  \frac{dS}{dt} & = - \beta I \frac{S}{N}, & \frac{dI}{dt} & = - \gamma I + \beta I \frac{S}{N}, & \frac{dR}{dt} & = \gamma I.
\end{align}
When fitting to qualitatively different data, the two dimensional SIR model manifold is compressed depending on the informativity of the available data.
The compression determines which parameters are identifiable from data and leads to an appropriate reduced model.

We focus on two reduced models on the boundary of the SIR model, shown in Fig.~\ref{SIR}b.
The first boundary segment, corresponding to $\gamma \rightarrow 0$, is the model with no recovery compartment, i.e., an ``SI'' model.
Similarly, the ``IR'' model with $\beta \rightarrow \infty$ has a very fast infection rate.
Consider only data from early times (red in Fig.~\ref{SIR}).
The FIM compresses the model manifold along the SI boundary segment, rendering the recovery rate $\gamma$ irrelevant.
The approximate SI model (red dashed line in Fig.~\ref{SIR}) has an effective infection rate that fits the early exponential growth
\footnote{Although this approximation is constructed by taking $\gamma \rightarrow 0$, it does not require the ``true'' value of $\gamma$ to be small.
Rather, the role of the recovery mechanism can be compressed into a simpler model with an effective infection rate, similar to the effective electron mass in a condensed matter system.}.
However, recovery data at later stages (blue), render $\beta$ irrelevant and are well approximated by the IR model.

\begin{figure*}
    \centering
    \includegraphics[width=7in]{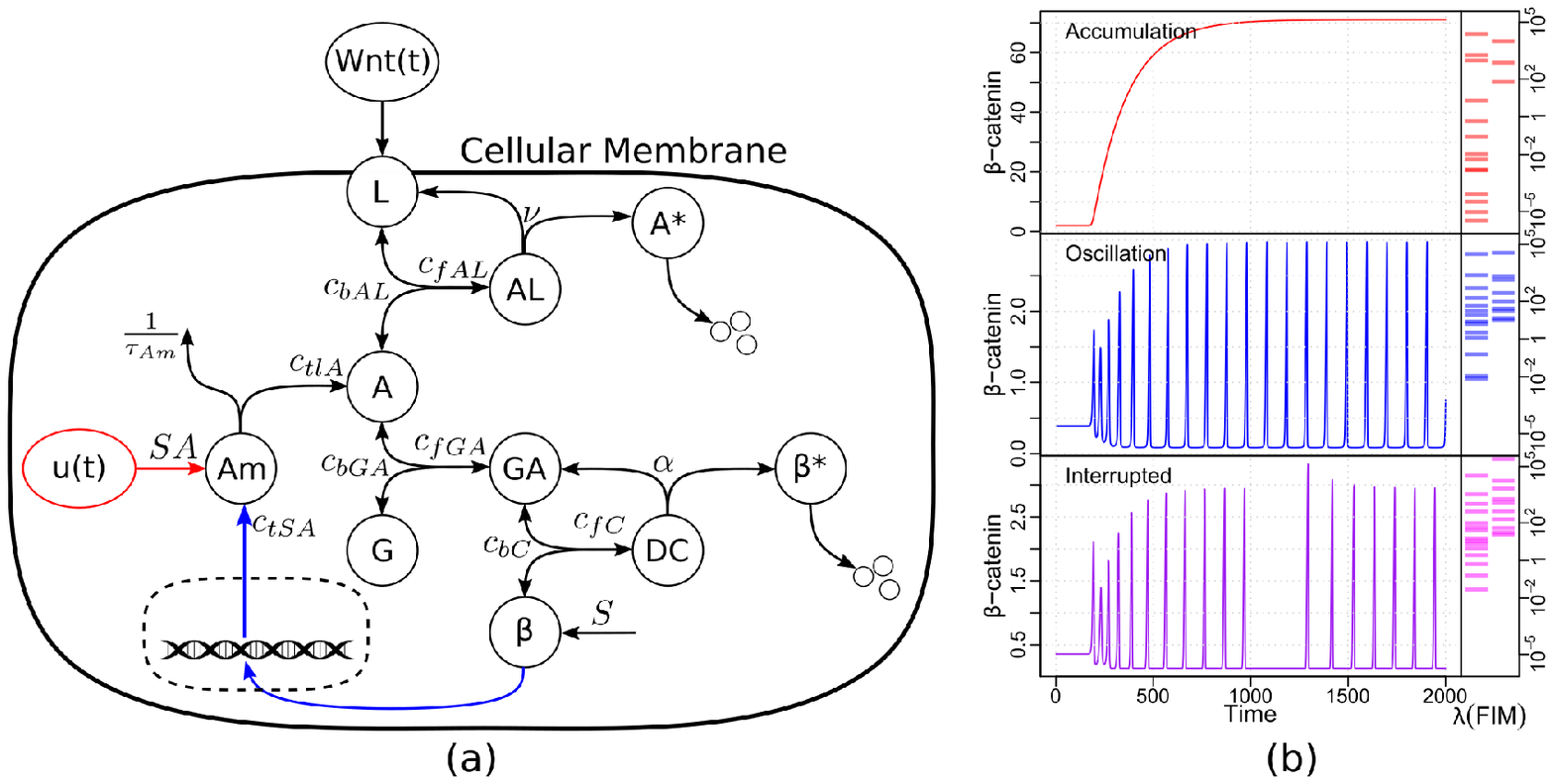}
    \caption{(a) A network diagram showing the ``full" model of the Wnt pathway.
    The red indicates mechanisms that are unique to the accumulation model, and blue indicates mechanisms unique to the oscillation model.
    (b) Three possible behaviors of beta-catenin in response to Wnt.
    The accumulation \cite{goentoro2009} and oscillation \cite{jensen2010} behaviors are known to occur naturally, while behaviors similar to the interrupted behavior have been reported in \cite{Riedel-Kruse2007, Gibb2009, Goldbeter2008, Gomez2008, Rui2007}.}
    \label{fig:wnt}
\end{figure*}

The SI and IR models interpolate in their respective domains, but fail to transfer beyond those domains.
The SIR model is the simplest that can interpolate all three regimes.
Formally, the hierarchy of potential models forms a graded Partially Ordered Set (POSet).
A POSet generalizes the concept of order within a set.
Real numbers are completely ordered, that is $\forall a,b \in \mathbb{R}$, with $a \neq b$, either $a>b$ or $a<b$.
POSets additionally allow for elements to be \emph{incomparable}, i.e., neither  $a > b$ nor $a < b$.
Discrete POSets can be represented by a directed graph known as a Hasse diagram \cite{transtrum2014infortopol} as in Fig.~\ref{SIR}c.
In this formalism, the SI and IR models are incomparable; there is no path in the directed graph connecting them.
The SIR model is the \emph{supremum} (i.e., least upper bound) of the SI and IR models as it is the simplest model connected to both the SI and IR models within the Hasse diagram.
The topological relationships (the adjacency relationships summarized in the Hasse diagram) among candidate models enable reasoning about the mechanisms at play in diverse contexts and inform the construction of the supremal model which minimally merges model elements.
The resulting supremal model is more expressive than either of its children, and so enables predictions under qualitatively different conditions than either training set.
This is because the supremal model's additional parameters have been identified by the reduction steps as meaningful, and so by definition must create at least one novel behavior in combination.

This simple example suggests the possibility of an algorithm for finding supremal models.
The next example will introduce mathematical concepts necessary for a general algorithm, but the conceptual steps in the process are already clear.
First, select a hypothesis space, i.e., pick a function form for a model that describes the behaviors of interest, the MSEIR model in this case.
Second, reduce the model via MBAM to find minimal models that described each behavior of interest, the IR and SI models.
Finally, find the reductions that are common to each of the child models and apply them to the full model.
In this case, the SI model removed all parameters except $\beta$ while the IR model removed all parameters except $\gamma$ so the supremal model is the SIR model, the simplest model to include both parameters.
In generic scenarios, some of the reductions may combine parameters in the reduced models in ways that obscure which are the common approximations.
The example below illustrates this possibility and introduces a formalism to deal with it.

The supremum principle is applicable to any hierarchical family of models.
In this paper, we focus specifically on hierarchies generated by MBAM, which includes things as diverse as power systems \cite{svenda2021state, saric2020symbolic, francis2019network, transtrum2016measurement}, systems biology \cite{jeong2018experimental, transtrum2016bridging, mannakee2016sloppiness, transtrum2015perspective}, materials science \cite{kurniawan2021bayesian}, biogeochemistry \cite{marschmann2019equifinality}, nuclear physics \cite{nikvsic2017sloppy}, neuroscience \cite{rasband2021two}, and others \cite{pare2019model, gerach2019observation, lombardo2017systematic, pare2015unified}.
To better illustrate the general algorithm, we demonstrate the construction of supremal models in the supplement with a simple network spin model, and in a more complex biological system below.

The Wnt signaling pathway induces cell division in animals, and is one of the best studied in all of biology (See Fig~\ref{fig:wnt}a and the Appendix).
Via a multi-step process, an extracellular Wnt molecule causes a change in intracellular levels of the transcription factor $\beta$-catenin.
\emph{In vivo,} $\beta$-catenin either ``accumulates'' to a new, higher equilibrium \cite{lee2003, goentoro2009}, or ``oscillates'' between a low baseline and periodic spikes of high concentration \cite{jensen2010}, as illustrated in Fig.~\ref{fig:wnt}b.

The hypothesis space, i.e., functional form, we chose to model these two phenomena is a slight adaptation of that proposed by Jensen et al. \cite{jensen2010}, summarized in Fig~\ref{fig:wnt}a.
We adapt this model to the accumulation phase by removing the negative feedback loop and replacing it with a controllable activation of Axin2 (as \emph{in vivo} by USP7 \cite{ji_usp7_2019} among others) denoted by $u(t)$ in Fig.~\ref{fig:wnt}.
Fig.~\ref{fig:wnt}b presents characteristic time series for each of these two models.
In each case, the system begins in steady state and a Wnt stimulus is introduced at 200 minutes.
In the first case, $\beta$-catenin accumulates and equilibrates at a new steady state \cite{lee2003, goentoro2009}.
In the second case, the negative feedback loop triggers a Hopf bifurcation leading to sustained oscillations \cite{jensen2010}.

\begin{figure}
\centering
\includegraphics[width=0.5\textwidth]{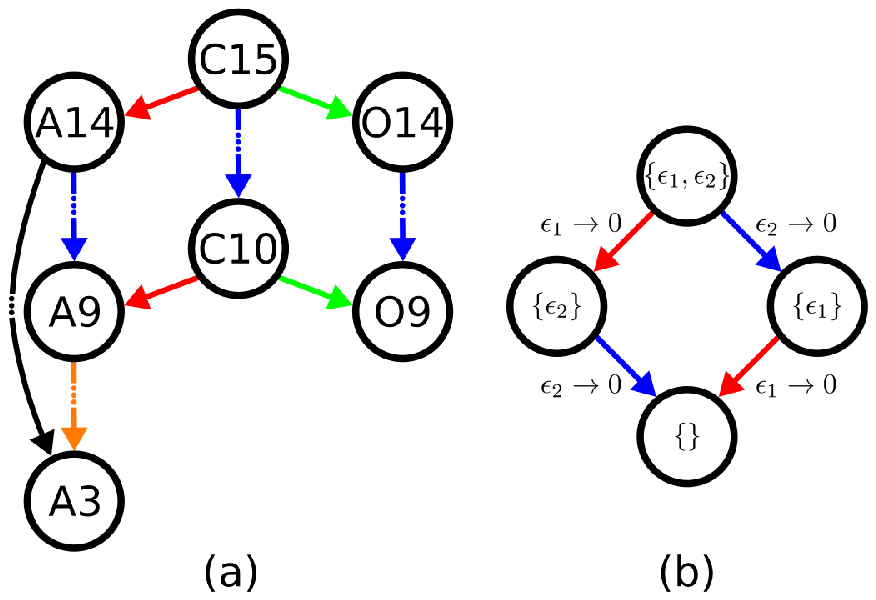}
\caption{(a) The Hasse diagram of key reduced models for the Wnt system.
The nodes (models) are labeled by ``A" for accumulation, ``O" for oscillation, ``C" for combined, as well as the number of parameters each model contains.
The black line represents the original sequence of reduced accumulation models and the blue line represents the new sequence with parameter limits reordered to match those in the oscillation models.
The central, blue arrow shows that the original model can be reduced to the supremal model using the same set of limits used to arrive at A9 and O9.
Details about the model equations and sequence of limits can be found in the supplemental material.
(b) A general example of the diamond property.
Consider a model with $N$ parameters containing two parameters $\epsilon_1$ and $\epsilon_2$.
The diamond property states that the order in which parameters can be removed commutes.
The same model with $N-2$ parameters can be reached by first taking either $\epsilon_1$ or $\epsilon_2$ to zero, and then taking the other to zero.
The diamond property is used to reorder the limits of a sequence, to build a supremal model.}
\label{fig:hasse_diamond}
\end{figure}

Each model has 14 parameters.
A sloppy model analysis \cite{gutenkunst_universally_2007, transtrum_geometry_2011} reveals many small eigenvalues (right panel in Fig.~\ref{fig:wnt}b) in the respective FIM, indicating that many parameters are unidentifiable.
We remove irrelevant parameters using the Manifold Boundary Approximation Method (MBAM) \cite{transtrum2014modelreduc}, as summarized in Fig.~\ref{fig:hasse_diamond}a.
The accumulation behavior is minimally described by three parameters while the oscillation phenomenon requires nine.

The MBAM reductions are not black boxes.
Although the reduced models do abstract away many specific details, they retain vestiges of the full mechanisms, analogous to the SI and IR models in our epidemiology example.
To relate these behaviors, we now seek the \emph{supremum} of these two minimal representations.
However, unlike the simple example, the minimal Wnt models contain partially overlapping combinations of parameters, so the construction is non-trivial.

Each reduction can be rewritten as a single parameter taken to zero.
For example, consider an equilibrium approximation, i.e., $c_f, c_b \rightarrow \infty$.
This can be rewritten as a time constant going to zero $\tau_f = 1/c_f \rightarrow 0$ and a nonzero equilibrium constant $K_D = c_b/c_f$.
This form, however, is not unique as we could also have chosen $\tau_b = 1/c_b \rightarrow 0$.
For all equilibrium approximations we adopt the first as a standard form.

Next, we observe that the same reduced models could be derived by applying the same approximations in different orders.
Commuting the order of reductions creates a diamond motif in the Hasse diagram, as in Fig.~\ref{fig:hasse_diamond}b.
Because of the ambiguity in how reductions are labeled, consecutive limits including the same parameters can obscure this commutation relation.
For example, consider the consecutive limits of an equilibrium approximation ($c_{bGA} \rightarrow \infty$, $c_{fGA} \rightarrow \infty$ with $c_{bGA}/c_{fGA}$ constant and finite) followed by an irreversible approximation ($c_{bGA}/c_{fGA} \rightarrow 0$, $c_{fAL} \rightarrow \infty$).
Reparameterizing as $\epsilon_1=1/c_{fGA}$, $\epsilon_2=1/c_{fAL}$, and $\phi=c_{fAL}c_{bGA}/c_{fGA}$ makes the diamond property apparent.
The two limits of the diamond property can now be written as $\epsilon_1 \rightarrow 0$ and $\epsilon_2 \rightarrow 0$.

Writing all of the reductions in a standard form allows us to identify the approximations common to both reduced models.
Applying these common approximations to the original full model constructs the supremal model, as illustrated by the blue line connecting C15 to C10 in Fig.~\ref{fig:hasse_diamond}.
This process motivates a general algorithm for finding supremal models, and is given below.
\begin{enumerate}
    \item Define the hypothesis space by selecting a complex, multiparameter model to describe all desired behaviors.
    \item Perform MBAM to find reduced models that minimally describe each behavior.
    \item Reparameterize the models to detangle any conflated limits and find the approximations common to the both reduced models.
    \item Apply those common approximations to the original, full model to obtain the supremal model.
\end{enumerate}

Each of these steps is illustrated in Fig.~\ref{fig:hasse_diamond}.
The original hypothesis space is represented by C15 (step 1).  
Models A3 and O9 minimally describe the accumulation and oscillation behaviors, respectively (step 2).
Using the diamond property, illustrated in Fig.~\ref{fig:hasse_diamond}b, we reparameterize and find the approximations common to each reduced model, represented by the blue arrows (step 3).
Applying these common approximations to C15 gives the supremal model, C10 (step 4).
This process is described in more detail in the supplementary material, including a discussion on fitting the supremal model parameters and application to the Wnt model.
Additionally, the supplemental material presents a second algorithm that exploits a general duality inherent in POSets.

Since the supremum has more parameters than either of the reduced models, the original data sets cannot individually constrain all of the supremal parameters.  
However, since each parameter is constrained by one data set of the other (e.g., $SA$ is constrained by the accumulation data but not eh oscillation data), fitting the supremal model to both data sets simultaneously does identify each parameter.
By including both the feedback loop and external control, and their associated parameters, the supremum enables the accumulation and oscillation phenomena, as well as additional behaviors neither A14 nor O14 can produce.
Fig.~\ref{fig:wnt}b demonstrates one such example, the ``interrupted" behavior, in which the external control modulates the phase of the oscillation.
Regular oscillatory behavior in the Wnt pathway is well-known \emph{in vivo} from the segmentation clock in vertebrate embryos along the anterior-posterior axis to establish, for example, the repeating pattern of vertebrae and ribs \cite{Pourquie2003}.
These regular oscillations can have their period and phase modified, stopped, or restarted through manipulation of ``dorsalizing'' or ``ventralizing'' molecular regulators, much like the interrupted behavior we see in the supremal model \cite{Riedel-Kruse2007, Gibb2009, Goldbeter2008, Gomez2008, Rui2007}.
To validate our model, we apply MBAM to the full model using data for the interrupted regime.
This reduction gives the supremal model constructed by our algorithm, indicating that the supremum is the model that would have been selected had observations been available for this behavior.

The supremum principle shows promise for transferring predictability to truly new domains.
For example, the SI and IR models fail in the intermediate regime, and the accumulation and oscillation models fail in the interrupted regime, but the supremal models in each case are able to embrace all three behaviors.
It does this by including key modeling elements (e.g., feedback and external control) that are missing from the reduced models.
Since the supremal model combines distinct modeling elements, it enables new behaviors in regimes in which those modeling elements are all necessary.
With a different starting model, couched in a different hypothesis space, the supremal model will be different, but it will still transfer according to the given hypothesis.
This is more than the simple generalization of, e.g., multi-task learning (MTL) \cite{Caruana1997}.
Supremal models apply in a more global way; they aim to improve the transferability to data in a completely new regime.

Classical psychological theories use geometric constructions to represent analogical relationships.
Most notably, in the parallelogram model \cite{rumelhart1973model}, an analogy such as man:king::woman:queen is represented as four corners of a parallelogram with analogical relationships forming parallel sides \cite{peterson2020parallelograms}.
Such constructions are widespread in AI applications ranging from recommender systems \cite{musto2010enhanced} to natural language processing \cite{reid2018vector}.
The key property, however, is the topological relationship between analogous elements \cite{Gentner1983} that for parallelograms form the same diamond motif as in Fig.~\ref{fig:hasse_diamond}.
The analogical relationships among words are the same as those between reduced models.
Kings are subsets of men just as reduced models are restricted cases of more general models, and classifications based on royalty analogize across genders just as approximations transfer across models.
Thus, the supremum construction identifies the mathematical ``analogies" between models by teasing out the common mechanisms or analogous reductions (see the colored arrows in Fig.~\ref{fig:hasse_diamond}).
The approximations in linking model C15 to model C10 are the same as those connecting model A14 to model A9, i.e., C15:C10::A14:A9.
The colored arrows indicate the many other possible analogies that could be drawn among the models.

The supremum principle is applicable to any hierarchical family of models, and so there are some inherent limitations and potential extensions.
First, the algorithm we present here is specific to hierarchies generated by MBAM, but future work could consider other families.
Next, the result depends on the hierarchy one uses, for example, our Wnt study used the hierarchy generated by the model of Jensen et al. \cite{jensen2010}.
Given different hierarchies, supremal models are a principled way of reasoning about the implications of those hypothesis.
Future work may use supremal models to guide experimental design for hypothesis testing or parameter estimation.
Finally, one could consider models that are derived independently of a hierarchical family.
Future work could explore how to most naturally embed such models within a hierarchy to enable transferability.

Beyond the appeal of elegant, simplified models, we expect supremal models to be of broad practical use; for example, in systems that need a controller to move between two behavioral states, but is difficult to fully model and a reduced model is needed.
Such systems include shifting from diseased to healthy states in medical contexts, failing to stable power grids in electrical engineering, ductile to brittle structures in material science, and collapsed to restored resources in ecosystem-based management. 
Supremal models are also designed for maximal simplicity while retaining some transferability, i.e., attempting to predict in regimes not yet examined, such as in climate modeling, prosperous non-growth-based economics, and human behavior during a pandemic.
Practitioners from a wide variety of fields will find supremum modeling a powerful addition to their toolboxes.

This work was supported by the US National Science Foundation under Award NSF-1753357 (CP, CA, MKT), CMMT-1834332 (CP, MKT), and EPCN-1710727 (CA, MKT).
We thank Sean Warnick, Kolten Barfuss, and Alex Stankovic for helpful conversations.
We thank Ben Francis, Dan Karls, Ellad Tadmor, and Ryan Elliott and two anonymous reviewers for comments on the manuscript.

\bibliography{refs}

\begin{thebibliography}{52}%
\makeatletter
\providecommand \@ifxundefined [1]{%
 \@ifx{#1\undefined}
}%
\providecommand \@ifnum [1]{%
 \ifnum #1\expandafter \@firstoftwo
 \else \expandafter \@secondoftwo
 \fi
}%
\providecommand \@ifx [1]{%
 \ifx #1\expandafter \@firstoftwo
 \else \expandafter \@secondoftwo
 \fi
}%
\providecommand \natexlab [1]{#1}%
\providecommand \enquote  [1]{``#1''}%
\providecommand \bibnamefont  [1]{#1}%
\providecommand \bibfnamefont [1]{#1}%
\providecommand \citenamefont [1]{#1}%
\providecommand \href@noop [0]{\@secondoftwo}%
\providecommand \href [0]{\begingroup \@sanitize@url \@href}%
\providecommand \@href[1]{\@@startlink{#1}\@@href}%
\providecommand \@@href[1]{\endgroup#1\@@endlink}%
\providecommand \@sanitize@url [0]{\catcode `\\12\catcode `\$12\catcode
  `\&12\catcode `\#12\catcode `\^12\catcode `\_12\catcode `\%12\relax}%
\providecommand \@@startlink[1]{}%
\providecommand \@@endlink[0]{}%
\providecommand \url  [0]{\begingroup\@sanitize@url \@url }%
\providecommand \@url [1]{\endgroup\@href {#1}{\urlprefix }}%
\providecommand \urlprefix  [0]{URL }%
\providecommand \Eprint [0]{\href }%
\providecommand \doibase [0]{http://dx.doi.org/}%
\providecommand \selectlanguage [0]{\@gobble}%
\providecommand \bibinfo  [0]{\@secondoftwo}%
\providecommand \bibfield  [0]{\@secondoftwo}%
\providecommand \translation [1]{[#1]}%
\providecommand \BibitemOpen [0]{}%
\providecommand \bibitemStop [0]{}%
\providecommand \bibitemNoStop [0]{.\EOS\space}%
\providecommand \EOS [0]{\spacefactor3000\relax}%
\providecommand \BibitemShut  [1]{\csname bibitem#1\endcsname}%
\let\auto@bib@innerbib\@empty
\bibitem [{\citenamefont {Baxter}(2000)}]{baxter2000model}%
  \BibitemOpen
  \bibfield  {author} {\bibinfo {author} {\bibfnamefont {J.}~\bibnamefont
  {Baxter}},\ }\href {\doibase 10.1613/jair.731} {\bibfield  {journal}
  {\bibinfo  {journal} {Journal of Artificial Intelligence Research}\ }\textbf
  {\bibinfo {volume} {12}},\ \bibinfo {pages} {149} (\bibinfo {year}
  {2000})}\BibitemShut {NoStop}%
\bibitem [{\citenamefont {Weiss}\ \emph {et~al.}(2016)\citenamefont {Weiss},
  \citenamefont {Khoshgoftaar},\ and\ \citenamefont {Wang}}]{weiss2016survey}%
  \BibitemOpen
  \bibfield  {author} {\bibinfo {author} {\bibfnamefont {K.}~\bibnamefont
  {Weiss}}, \bibinfo {author} {\bibfnamefont {T.~M.}\ \bibnamefont
  {Khoshgoftaar}}, \ and\ \bibinfo {author} {\bibfnamefont {D.}~\bibnamefont
  {Wang}},\ }\href {\doibase 10.1186/s40537-016-0043-6} {\bibfield  {journal}
  {\bibinfo  {journal} {Journal of Big Data}\ }\textbf {\bibinfo {volume}
  {3}},\ \bibinfo {pages} {9} (\bibinfo {year} {2016})}\BibitemShut {NoStop}%
\bibitem [{\citenamefont {Batterman}()}]{batterman2001devil}%
  \BibitemOpen
  \bibfield  {author} {\bibinfo {author} {\bibfnamefont {R.~W.}\ \bibnamefont
  {Batterman}},\ }\href@noop {} {\emph {\bibinfo {title} {The {Devil} in the
  {Details}: {Asymptotic} {Reasoning} in {Explanation}, {Reduction}, and
  {Emergence}}}}\ (\bibinfo  {publisher} {Oxford University Press})\ \bibinfo
  {note} {google-Books-ID: EiIM5koj\_J0C}\BibitemShut {NoStop}%
\bibitem [{\citenamefont {Brown}\ and\ \citenamefont
  {Sethna}(2003)}]{brown2003statismechan}%
  \BibitemOpen
  \bibfield  {author} {\bibinfo {author} {\bibfnamefont {K.~S.}\ \bibnamefont
  {Brown}}\ and\ \bibinfo {author} {\bibfnamefont {J.~P.}\ \bibnamefont
  {Sethna}},\ }\href {\doibase 10.1103/physreve.68.021904} {\bibfield
  {journal} {\bibinfo  {journal} {Physical Review E}\ }\textbf {\bibinfo
  {volume} {68}},\ \bibinfo {pages} {021904} (\bibinfo {year}
  {2003})}\BibitemShut {NoStop}%
\bibitem [{\citenamefont {Brown}\ \emph {et~al.}(2004)\citenamefont {Brown},
  \citenamefont {Hill}, \citenamefont {Calero}, \citenamefont {Myers},
  \citenamefont {Lee}, \citenamefont {Sethna},\ and\ \citenamefont
  {Cerione}}]{brown2004statismechan}%
  \BibitemOpen
  \bibfield  {author} {\bibinfo {author} {\bibfnamefont {K.~S.}\ \bibnamefont
  {Brown}}, \bibinfo {author} {\bibfnamefont {C.~C.}\ \bibnamefont {Hill}},
  \bibinfo {author} {\bibfnamefont {G.~A.}\ \bibnamefont {Calero}}, \bibinfo
  {author} {\bibfnamefont {C.~R.}\ \bibnamefont {Myers}}, \bibinfo {author}
  {\bibfnamefont {K.~H.}\ \bibnamefont {Lee}}, \bibinfo {author} {\bibfnamefont
  {J.~P.}\ \bibnamefont {Sethna}}, \ and\ \bibinfo {author} {\bibfnamefont
  {R.~A.}\ \bibnamefont {Cerione}},\ }\href {\doibase
  10.1088/1478-3967/1/3/006} {\bibfield  {journal} {\bibinfo  {journal}
  {Physical Biology}\ }\textbf {\bibinfo {volume} {1}},\ \bibinfo {pages} {184}
  (\bibinfo {year} {2004})}\BibitemShut {NoStop}%
\bibitem [{\citenamefont {Waterfall}\ \emph {et~al.}(2006)\citenamefont
  {Waterfall}, \citenamefont {Casey}, \citenamefont {Gutenkunst}, \citenamefont
  {Brown}, \citenamefont {Myers}, \citenamefont {Brouwer}, \citenamefont
  {Elser},\ and\ \citenamefont {Sethna}}]{waterfall2006sloppmodel}%
  \BibitemOpen
  \bibfield  {author} {\bibinfo {author} {\bibfnamefont {J.~J.}\ \bibnamefont
  {Waterfall}}, \bibinfo {author} {\bibfnamefont {F.~P.}\ \bibnamefont
  {Casey}}, \bibinfo {author} {\bibfnamefont {R.~N.}\ \bibnamefont
  {Gutenkunst}}, \bibinfo {author} {\bibfnamefont {K.~S.}\ \bibnamefont
  {Brown}}, \bibinfo {author} {\bibfnamefont {C.~R.}\ \bibnamefont {Myers}},
  \bibinfo {author} {\bibfnamefont {P.~W.}\ \bibnamefont {Brouwer}}, \bibinfo
  {author} {\bibfnamefont {V.}~\bibnamefont {Elser}}, \ and\ \bibinfo {author}
  {\bibfnamefont {J.~P.}\ \bibnamefont {Sethna}},\ }\href {\doibase
  10.1103/physrevlett.97.150601} {\bibfield  {journal} {\bibinfo  {journal}
  {Physical Review Letters}\ }\textbf {\bibinfo {volume} {97}},\ \bibinfo
  {pages} {150601} (\bibinfo {year} {2006})}\BibitemShut {NoStop}%
\bibitem [{\citenamefont {Machta}\ \emph {et~al.}(2013)\citenamefont {Machta},
  \citenamefont {Chachra}, \citenamefont {Transtrum},\ and\ \citenamefont
  {Sethna}}]{machta2013paramspace}%
  \BibitemOpen
  \bibfield  {author} {\bibinfo {author} {\bibfnamefont {B.~B.}\ \bibnamefont
  {Machta}}, \bibinfo {author} {\bibfnamefont {R.}~\bibnamefont {Chachra}},
  \bibinfo {author} {\bibfnamefont {M.~K.}\ \bibnamefont {Transtrum}}, \ and\
  \bibinfo {author} {\bibfnamefont {J.~P.}\ \bibnamefont {Sethna}},\ }\href
  {\doibase 10.1126/science.1238723} {\bibfield  {journal} {\bibinfo  {journal}
  {Science}\ }\textbf {\bibinfo {volume} {342}},\ \bibinfo {pages} {604}
  (\bibinfo {year} {2013})}\BibitemShut {NoStop}%
\bibitem [{\citenamefont {Amari}\ and\ \citenamefont
  {Nagaoka}(2007)}]{amari2007methods}%
  \BibitemOpen
  \bibfield  {author} {\bibinfo {author} {\bibfnamefont {S.-i.}\ \bibnamefont
  {Amari}}\ and\ \bibinfo {author} {\bibfnamefont {H.}~\bibnamefont
  {Nagaoka}},\ }\href@noop {} {\emph {\bibinfo {title} {Methods of information
  geometry}}},\ Vol.\ \bibinfo {volume} {191}\ (\bibinfo  {publisher} {American
  Mathematical Soc.},\ \bibinfo {year} {2007})\BibitemShut {NoStop}%
\bibitem [{\citenamefont {Transtrum}\ \emph {et~al.}(2010)\citenamefont
  {Transtrum}, \citenamefont {Machta},\ and\ \citenamefont
  {Sethna}}]{transtrum2010}%
  \BibitemOpen
  \bibfield  {author} {\bibinfo {author} {\bibfnamefont {M.~K.}\ \bibnamefont
  {Transtrum}}, \bibinfo {author} {\bibfnamefont {B.~B.}\ \bibnamefont
  {Machta}}, \ and\ \bibinfo {author} {\bibfnamefont {J.~P.}\ \bibnamefont
  {Sethna}},\ }\href {\doibase 10.1103/PhysRevLett.104.060201} {\bibfield
  {journal} {\bibinfo  {journal} {Phys. Rev. Lett.}\ }\textbf {\bibinfo
  {volume} {104}},\ \bibinfo {pages} {060201} (\bibinfo {year}
  {2010})}\BibitemShut {NoStop}%
\bibitem [{\citenamefont {LaMont}\ and\ \citenamefont
  {Wiggins}(2019)}]{lamont2019correspondence}%
  \BibitemOpen
  \bibfield  {author} {\bibinfo {author} {\bibfnamefont {C.~H.}\ \bibnamefont
  {LaMont}}\ and\ \bibinfo {author} {\bibfnamefont {P.~A.}\ \bibnamefont
  {Wiggins}},\ }\href {\doibase 10.1103/PhysRevE.99.052140} {\bibfield
  {journal} {\bibinfo  {journal} {Physical Review E}\ }\textbf {\bibinfo
  {volume} {99}},\ \bibinfo {pages} {052140} (\bibinfo {year}
  {2019})}\BibitemShut {NoStop}%
\bibitem [{\citenamefont {Quinn}\ \emph {et~al.}(2019)\citenamefont {Quinn},
  \citenamefont {Wilber}, \citenamefont {Townsend},\ and\ \citenamefont
  {Sethna}}]{quinn2019}%
  \BibitemOpen
  \bibfield  {author} {\bibinfo {author} {\bibfnamefont {K.~N.}\ \bibnamefont
  {Quinn}}, \bibinfo {author} {\bibfnamefont {H.}~\bibnamefont {Wilber}},
  \bibinfo {author} {\bibfnamefont {A.}~\bibnamefont {Townsend}}, \ and\
  \bibinfo {author} {\bibfnamefont {J.~P.}\ \bibnamefont {Sethna}},\ }\href
  {\doibase 10.1103/PhysRevLett.122.158302} {\bibfield  {journal} {\bibinfo
  {journal} {Physical Review Letters}\ }\textbf {\bibinfo {volume} {122}},\
  \bibinfo {pages} {158302} (\bibinfo {year} {2019})}\BibitemShut {NoStop}%
\bibitem [{\citenamefont {Transtrum}\ \emph
  {et~al.}(2011{\natexlab{a}})\citenamefont {Transtrum}, \citenamefont
  {Machta},\ and\ \citenamefont {Sethna}}]{transtrum2011geometnonlin}%
  \BibitemOpen
  \bibfield  {author} {\bibinfo {author} {\bibfnamefont {M.~K.}\ \bibnamefont
  {Transtrum}}, \bibinfo {author} {\bibfnamefont {B.~B.}\ \bibnamefont
  {Machta}}, \ and\ \bibinfo {author} {\bibfnamefont {J.~P.}\ \bibnamefont
  {Sethna}},\ }\href {\doibase 10.1103/physreve.83.036701} {\bibfield
  {journal} {\bibinfo  {journal} {Physical Review E}\ }\textbf {\bibinfo
  {volume} {83}},\ \bibinfo {pages} {036701} (\bibinfo {year}
  {2011}{\natexlab{a}})}\BibitemShut {NoStop}%
\bibitem [{\citenamefont {Lake}\ \emph {et~al.}(2017)\citenamefont {Lake},
  \citenamefont {Ullman}, \citenamefont {Tenenbaum},\ and\ \citenamefont
  {Gershman}}]{lake2017building}%
  \BibitemOpen
  \bibfield  {author} {\bibinfo {author} {\bibfnamefont {B.~M.}\ \bibnamefont
  {Lake}}, \bibinfo {author} {\bibfnamefont {T.~D.}\ \bibnamefont {Ullman}},
  \bibinfo {author} {\bibfnamefont {J.~B.}\ \bibnamefont {Tenenbaum}}, \ and\
  \bibinfo {author} {\bibfnamefont {S.~J.}\ \bibnamefont {Gershman}},\ }\href
  {\doibase 10.1017/S0140525X16001837} {\bibfield  {journal} {\bibinfo
  {journal} {Behavioral and Brain Sciences}\ }\textbf {\bibinfo {volume}
  {40}},\ \bibinfo {pages} {e253} (\bibinfo {year} {2017})}\BibitemShut
  {NoStop}%
\bibitem [{\citenamefont {Webb}\ \emph {et~al.}(2020)\citenamefont {Webb},
  \citenamefont {Dulberg}, \citenamefont {Frankland}, \citenamefont {Petrov},
  \citenamefont {O'Reilly},\ and\ \citenamefont {Cohen}}]{webb2020learning}%
  \BibitemOpen
  \bibfield  {author} {\bibinfo {author} {\bibfnamefont {T.}~\bibnamefont
  {Webb}}, \bibinfo {author} {\bibfnamefont {Z.}~\bibnamefont {Dulberg}},
  \bibinfo {author} {\bibfnamefont {S.}~\bibnamefont {Frankland}}, \bibinfo
  {author} {\bibfnamefont {A.}~\bibnamefont {Petrov}}, \bibinfo {author}
  {\bibfnamefont {R.}~\bibnamefont {O'Reilly}}, \ and\ \bibinfo {author}
  {\bibfnamefont {J.}~\bibnamefont {Cohen}},\ }in\ \href
  {http://proceedings.mlr.press/v119/webb20a.html} {\emph {\bibinfo {booktitle}
  {Proceedings of the 37th International Conference on Machine Learning}}},\
  \bibinfo {series} {Proceedings of Machine Learning Research}, Vol.\ \bibinfo
  {volume} {119},\ \bibinfo {editor} {edited by\ \bibinfo {editor}
  {\bibfnamefont {H.~D.}\ \bibnamefont {III}}\ and\ \bibinfo {editor}
  {\bibfnamefont {A.}~\bibnamefont {Singh}}}\ (\bibinfo  {publisher} {PMLR},\
  \bibinfo {year} {2020})\ pp.\ \bibinfo {pages} {10136--10146}\BibitemShut
  {NoStop}%
\bibitem [{\citenamefont {Brouwer}\ and\ \citenamefont
  {Eisenberg}(2018)}]{brouwer2018underlying}%
  \BibitemOpen
  \bibfield  {author} {\bibinfo {author} {\bibfnamefont {A.~F.}\ \bibnamefont
  {Brouwer}}\ and\ \bibinfo {author} {\bibfnamefont {M.~C.}\ \bibnamefont
  {Eisenberg}},\ }\href {http://arxiv.org/abs/1802.05641} {\bibfield  {journal}
  {\bibinfo  {journal} {arXiv:1802.05641 [math]}\ } (\bibinfo {year} {2018})},\
  \bibinfo {note} {arXiv: 1802.05641}\BibitemShut {NoStop}%
\bibitem [{\citenamefont {Transtrum}\ and\ \citenamefont
  {Qiu}(2014)}]{transtrum2014modelreduc}%
  \BibitemOpen
  \bibfield  {author} {\bibinfo {author} {\bibfnamefont {M.~K.}\ \bibnamefont
  {Transtrum}}\ and\ \bibinfo {author} {\bibfnamefont {P.}~\bibnamefont
  {Qiu}},\ }\href {\doibase 10.1103/physrevlett.113.098701} {\bibfield
  {journal} {\bibinfo  {journal} {Physical Review Letters}\ }\textbf {\bibinfo
  {volume} {113}},\ \bibinfo {pages} {098701} (\bibinfo {year}
  {2014})}\BibitemShut {NoStop}%
\bibitem [{\citenamefont {Transtrum}\ \emph {et~al.}(2015)\citenamefont
  {Transtrum}, \citenamefont {Machta}, \citenamefont {Brown}, \citenamefont
  {Daniels}, \citenamefont {Myers},\ and\ \citenamefont
  {Sethna}}]{transtrum2015perspective}%
  \BibitemOpen
  \bibfield  {author} {\bibinfo {author} {\bibfnamefont {M.~K.}\ \bibnamefont
  {Transtrum}}, \bibinfo {author} {\bibfnamefont {B.~B.}\ \bibnamefont
  {Machta}}, \bibinfo {author} {\bibfnamefont {K.~S.}\ \bibnamefont {Brown}},
  \bibinfo {author} {\bibfnamefont {B.~C.}\ \bibnamefont {Daniels}}, \bibinfo
  {author} {\bibfnamefont {C.~R.}\ \bibnamefont {Myers}}, \ and\ \bibinfo
  {author} {\bibfnamefont {J.~P.}\ \bibnamefont {Sethna}},\ }\href@noop {}
  {\bibfield  {journal} {\bibinfo  {journal} {The Journal of chemical physics}\
  }\textbf {\bibinfo {volume} {143}},\ \bibinfo {pages} {07B201\_1} (\bibinfo
  {year} {2015})}\BibitemShut {NoStop}%
\bibitem [{Note1()}]{Note1}%
  \BibitemOpen
  \bibinfo {note} {Although this approximation is constructed by taking $\gamma
  \rightarrow 0$, it does not require the ``true'' value of $\gamma $ to be
  small. Rather, the role of the recovery mechanism can be compressed into a
  simpler model with an effective infection rate, similar to the effective
  electron mass in a condensed matter system.}\BibitemShut {Stop}%
\bibitem [{\citenamefont {Goentoro}\ and\ \citenamefont
  {Kirschner}(2009)}]{goentoro2009}%
  \BibitemOpen
  \bibfield  {author} {\bibinfo {author} {\bibfnamefont {L.}~\bibnamefont
  {Goentoro}}\ and\ \bibinfo {author} {\bibfnamefont {M.~W.}\ \bibnamefont
  {Kirschner}},\ }\href {\doibase 10.1016/j.molcel.2009.11.017} {\bibfield
  {journal} {\bibinfo  {journal} {Molecular Cell}\ }\textbf {\bibinfo {volume}
  {36}},\ \bibinfo {pages} {872} (\bibinfo {year} {2009})}\BibitemShut
  {NoStop}%
\bibitem [{\citenamefont {Jensen}\ \emph {et~al.}(2010)\citenamefont {Jensen},
  \citenamefont {Pedersen}, \citenamefont {Krishna},\ and\ \citenamefont
  {Jensen}}]{jensen2010}%
  \BibitemOpen
  \bibfield  {author} {\bibinfo {author} {\bibfnamefont {P.~B.}\ \bibnamefont
  {Jensen}}, \bibinfo {author} {\bibfnamefont {L.}~\bibnamefont {Pedersen}},
  \bibinfo {author} {\bibfnamefont {S.}~\bibnamefont {Krishna}}, \ and\
  \bibinfo {author} {\bibfnamefont {M.~H.}\ \bibnamefont {Jensen}},\ }\href
  {\doibase 10.1016/j.bpj.2009.11.039} {\bibfield  {journal} {\bibinfo
  {journal} {Biophysical Journal}\ }\textbf {\bibinfo {volume} {98}},\ \bibinfo
  {pages} {943} (\bibinfo {year} {2010})}\BibitemShut {NoStop}%
\bibitem [{\citenamefont {Riedel-Kruse}\ \emph {et~al.}(2007)\citenamefont
  {Riedel-Kruse}, \citenamefont {M{\"{u}}ller},\ and\ \citenamefont
  {Oates}}]{Riedel-Kruse2007}%
  \BibitemOpen
  \bibfield  {author} {\bibinfo {author} {\bibfnamefont {I.~H.}\ \bibnamefont
  {Riedel-Kruse}}, \bibinfo {author} {\bibfnamefont {C.}~\bibnamefont
  {M{\"{u}}ller}}, \ and\ \bibinfo {author} {\bibfnamefont {A.~C.}\
  \bibnamefont {Oates}},\ }\href {\doibase 10.1126/science.1142538} {\bibfield
  {journal} {\bibinfo  {journal} {Science}\ }\textbf {\bibinfo {volume}
  {317}},\ \bibinfo {pages} {1911} (\bibinfo {year} {2007})}\BibitemShut
  {NoStop}%
\bibitem [{\citenamefont {Gibb}\ \emph {et~al.}(2009)\citenamefont {Gibb},
  \citenamefont {Zagorska}, \citenamefont {Melton}, \citenamefont {Tenin},
  \citenamefont {Vacca}, \citenamefont {Trainor}, \citenamefont {Maroto},\ and\
  \citenamefont {Dale}}]{Gibb2009}%
  \BibitemOpen
  \bibfield  {author} {\bibinfo {author} {\bibfnamefont {S.}~\bibnamefont
  {Gibb}}, \bibinfo {author} {\bibfnamefont {A.}~\bibnamefont {Zagorska}},
  \bibinfo {author} {\bibfnamefont {K.}~\bibnamefont {Melton}}, \bibinfo
  {author} {\bibfnamefont {G.}~\bibnamefont {Tenin}}, \bibinfo {author}
  {\bibfnamefont {I.}~\bibnamefont {Vacca}}, \bibinfo {author} {\bibfnamefont
  {P.}~\bibnamefont {Trainor}}, \bibinfo {author} {\bibfnamefont
  {M.}~\bibnamefont {Maroto}}, \ and\ \bibinfo {author} {\bibfnamefont {J.~K.}\
  \bibnamefont {Dale}},\ }\href {\doibase 10.1016/j.ydbio.2009.02.035}
  {\bibfield  {journal} {\bibinfo  {journal} {Developmental Biology}\ }\textbf
  {\bibinfo {volume} {330}},\ \bibinfo {pages} {21} (\bibinfo {year}
  {2009})}\BibitemShut {NoStop}%
\bibitem [{\citenamefont {Goldbeter}\ and\ \citenamefont
  {Pourqui{\'{e}}}(2008)}]{Goldbeter2008}%
  \BibitemOpen
  \bibfield  {author} {\bibinfo {author} {\bibfnamefont {A.}~\bibnamefont
  {Goldbeter}}\ and\ \bibinfo {author} {\bibfnamefont {O.}~\bibnamefont
  {Pourqui{\'{e}}}},\ }\href {\doibase 10.1016/j.jtbi.2008.01.006} {\bibfield
  {journal} {\bibinfo  {journal} {Journal of Theoretical Biology}\ }\textbf
  {\bibinfo {volume} {252}},\ \bibinfo {pages} {574} (\bibinfo {year}
  {2008})}\BibitemShut {NoStop}%
\bibitem [{\citenamefont {Gomez}\ \emph {et~al.}(2008)\citenamefont {Gomez},
  \citenamefont {{\"{O}}zbudak}, \citenamefont {Wunderlich}, \citenamefont
  {Baumann}, \citenamefont {Lewis},\ and\ \citenamefont
  {Pourqui{\'{e}}}}]{Gomez2008}%
  \BibitemOpen
  \bibfield  {author} {\bibinfo {author} {\bibfnamefont {C.}~\bibnamefont
  {Gomez}}, \bibinfo {author} {\bibfnamefont {E.~M.}\ \bibnamefont
  {{\"{O}}zbudak}}, \bibinfo {author} {\bibfnamefont {J.}~\bibnamefont
  {Wunderlich}}, \bibinfo {author} {\bibfnamefont {D.}~\bibnamefont {Baumann}},
  \bibinfo {author} {\bibfnamefont {J.}~\bibnamefont {Lewis}}, \ and\ \bibinfo
  {author} {\bibfnamefont {O.}~\bibnamefont {Pourqui{\'{e}}}},\ }\href
  {\doibase 10.1038/nature07020} {\bibfield  {journal} {\bibinfo  {journal}
  {Nature}\ }\textbf {\bibinfo {volume} {454}},\ \bibinfo {pages} {335}
  (\bibinfo {year} {2008})}\BibitemShut {NoStop}%
\bibitem [{\citenamefont {Rui}\ \emph {et~al.}(2007)\citenamefont {Rui},
  \citenamefont {Xu}, \citenamefont {Xiong}, \citenamefont {Cao}, \citenamefont
  {Lin}, \citenamefont {Zhang}, \citenamefont {Chan}, \citenamefont {Luo},
  \citenamefont {Han}, \citenamefont {Lu}, \citenamefont {Ye}, \citenamefont
  {Zhou}, \citenamefont {Han}, \citenamefont {Meng},\ and\ \citenamefont
  {Lin}}]{Rui2007}%
  \BibitemOpen
  \bibfield  {author} {\bibinfo {author} {\bibfnamefont {Y.}~\bibnamefont
  {Rui}}, \bibinfo {author} {\bibfnamefont {Z.}~\bibnamefont {Xu}}, \bibinfo
  {author} {\bibfnamefont {B.}~\bibnamefont {Xiong}}, \bibinfo {author}
  {\bibfnamefont {Y.}~\bibnamefont {Cao}}, \bibinfo {author} {\bibfnamefont
  {S.}~\bibnamefont {Lin}}, \bibinfo {author} {\bibfnamefont {M.}~\bibnamefont
  {Zhang}}, \bibinfo {author} {\bibfnamefont {S.~C.}\ \bibnamefont {Chan}},
  \bibinfo {author} {\bibfnamefont {W.}~\bibnamefont {Luo}}, \bibinfo {author}
  {\bibfnamefont {Y.}~\bibnamefont {Han}}, \bibinfo {author} {\bibfnamefont
  {Z.}~\bibnamefont {Lu}}, \bibinfo {author} {\bibfnamefont {Z.}~\bibnamefont
  {Ye}}, \bibinfo {author} {\bibfnamefont {H.~M.}\ \bibnamefont {Zhou}},
  \bibinfo {author} {\bibfnamefont {J.}~\bibnamefont {Han}}, \bibinfo {author}
  {\bibfnamefont {A.}~\bibnamefont {Meng}}, \ and\ \bibinfo {author}
  {\bibfnamefont {S.~C.}\ \bibnamefont {Lin}},\ }\href {\doibase
  10.1016/j.devcel.2007.07.006} {\bibfield  {journal} {\bibinfo  {journal}
  {Developmental Cell}\ }\textbf {\bibinfo {volume} {13}},\ \bibinfo {pages}
  {268} (\bibinfo {year} {2007})}\BibitemShut {NoStop}%
\bibitem [{\citenamefont {Transtrum}\ \emph {et~al.}(2014)\citenamefont
  {Transtrum}, \citenamefont {Hart},\ and\ \citenamefont
  {Qiu}}]{transtrum2014infortopol}%
  \BibitemOpen
  \bibfield  {author} {\bibinfo {author} {\bibfnamefont {M.~K.}\ \bibnamefont
  {Transtrum}}, \bibinfo {author} {\bibfnamefont {G.}~\bibnamefont {Hart}}, \
  and\ \bibinfo {author} {\bibfnamefont {P.}~\bibnamefont {Qiu}},\ }\href
  {http://arxiv.org/abs/1409.6203v2} {\bibfield  {journal} {\bibinfo  {journal}
  {CoRR}\ } (\bibinfo {year} {2014})},\ \Eprint
  {http://arxiv.org/abs/1409.6203} {arXiv:1409.6203 [physics.data-an]}
  \BibitemShut {NoStop}%
\bibitem [{\citenamefont {Svenda}\ \emph {et~al.}(2021)\citenamefont {Svenda},
  \citenamefont {Transtrum}, \citenamefont {Francis}, \citenamefont {Saric},\
  and\ \citenamefont {Stankovic}}]{svenda2021state}%
  \BibitemOpen
  \bibfield  {author} {\bibinfo {author} {\bibfnamefont {V.~G.}\ \bibnamefont
  {Svenda}}, \bibinfo {author} {\bibfnamefont {M.~K.}\ \bibnamefont
  {Transtrum}}, \bibinfo {author} {\bibfnamefont {B.~L.}\ \bibnamefont
  {Francis}}, \bibinfo {author} {\bibfnamefont {A.~T.}\ \bibnamefont {Saric}},
  \ and\ \bibinfo {author} {\bibfnamefont {A.~M.}\ \bibnamefont {Stankovic}},\
  }\href@noop {} {\bibfield  {journal} {\bibinfo  {journal} {IEEE Transactions
  on Power Systems}\ } (\bibinfo {year} {2021})}\BibitemShut {NoStop}%
\bibitem [{\citenamefont {Sari{\'c}}\ \emph {et~al.}(2020)\citenamefont
  {Sari{\'c}}, \citenamefont {Sari{\'c}}, \citenamefont {Transtrum},\ and\
  \citenamefont {Stankovi{\'c}}}]{saric2020symbolic}%
  \BibitemOpen
  \bibfield  {author} {\bibinfo {author} {\bibfnamefont {A.~T.}\ \bibnamefont
  {Sari{\'c}}}, \bibinfo {author} {\bibfnamefont {A.~A.}\ \bibnamefont
  {Sari{\'c}}}, \bibinfo {author} {\bibfnamefont {M.~K.}\ \bibnamefont
  {Transtrum}}, \ and\ \bibinfo {author} {\bibfnamefont {A.~M.}\ \bibnamefont
  {Stankovi{\'c}}},\ }\href@noop {} {\bibfield  {journal} {\bibinfo  {journal}
  {IEEE Transactions on Power Systems}\ }\textbf {\bibinfo {volume} {36}},\
  \bibinfo {pages} {2390} (\bibinfo {year} {2020})}\BibitemShut {NoStop}%
\bibitem [{\citenamefont {Francis}\ \emph {et~al.}(2019)\citenamefont
  {Francis}, \citenamefont {Nuttall}, \citenamefont {Transtrum}, \citenamefont
  {Sari{\'c}},\ and\ \citenamefont {Stankovi{\'c}}}]{francis2019network}%
  \BibitemOpen
  \bibfield  {author} {\bibinfo {author} {\bibfnamefont {B.~L.}\ \bibnamefont
  {Francis}}, \bibinfo {author} {\bibfnamefont {J.~R.}\ \bibnamefont
  {Nuttall}}, \bibinfo {author} {\bibfnamefont {M.~K.}\ \bibnamefont
  {Transtrum}}, \bibinfo {author} {\bibfnamefont {A.~T.}\ \bibnamefont
  {Sari{\'c}}}, \ and\ \bibinfo {author} {\bibfnamefont {A.~M.}\ \bibnamefont
  {Stankovi{\'c}}},\ }in\ \href@noop {} {\emph {\bibinfo {booktitle} {2019
  North American Power Symposium (NAPS)}}}\ (\bibinfo {organization} {IEEE},\
  \bibinfo {year} {2019})\ pp.\ \bibinfo {pages} {1--6}\BibitemShut {NoStop}%
\bibitem [{\citenamefont {Transtrum}\ \emph {et~al.}(2016)\citenamefont
  {Transtrum}, \citenamefont {Sari{\'c}},\ and\ \citenamefont
  {Stankovi{\'c}}}]{transtrum2016measurement}%
  \BibitemOpen
  \bibfield  {author} {\bibinfo {author} {\bibfnamefont {M.~K.}\ \bibnamefont
  {Transtrum}}, \bibinfo {author} {\bibfnamefont {A.~T.}\ \bibnamefont
  {Sari{\'c}}}, \ and\ \bibinfo {author} {\bibfnamefont {A.~M.}\ \bibnamefont
  {Stankovi{\'c}}},\ }\href@noop {} {\bibfield  {journal} {\bibinfo  {journal}
  {IEEE Transactions on Power Systems}\ }\textbf {\bibinfo {volume} {32}},\
  \bibinfo {pages} {2243} (\bibinfo {year} {2016})}\BibitemShut {NoStop}%
\bibitem [{\citenamefont {Jeong}\ \emph {et~al.}(2018)\citenamefont {Jeong},
  \citenamefont {Zhuang}, \citenamefont {Transtrum}, \citenamefont {Zhou},\
  and\ \citenamefont {Qiu}}]{jeong2018experimental}%
  \BibitemOpen
  \bibfield  {author} {\bibinfo {author} {\bibfnamefont {J.~E.}\ \bibnamefont
  {Jeong}}, \bibinfo {author} {\bibfnamefont {Q.}~\bibnamefont {Zhuang}},
  \bibinfo {author} {\bibfnamefont {M.~K.}\ \bibnamefont {Transtrum}}, \bibinfo
  {author} {\bibfnamefont {E.}~\bibnamefont {Zhou}}, \ and\ \bibinfo {author}
  {\bibfnamefont {P.}~\bibnamefont {Qiu}},\ }\href@noop {} {\bibfield
  {journal} {\bibinfo  {journal} {Quantitative Biology}\ }\textbf {\bibinfo
  {volume} {6}},\ \bibinfo {pages} {287} (\bibinfo {year} {2018})}\BibitemShut
  {NoStop}%
\bibitem [{\citenamefont {Transtrum}\ and\ \citenamefont
  {Qiu}(2016)}]{transtrum2016bridging}%
  \BibitemOpen
  \bibfield  {author} {\bibinfo {author} {\bibfnamefont {M.~K.}\ \bibnamefont
  {Transtrum}}\ and\ \bibinfo {author} {\bibfnamefont {P.}~\bibnamefont
  {Qiu}},\ }\href@noop {} {\bibfield  {journal} {\bibinfo  {journal} {PLoS
  computational biology}\ }\textbf {\bibinfo {volume} {12}},\ \bibinfo {pages}
  {e1004915} (\bibinfo {year} {2016})}\BibitemShut {NoStop}%
\bibitem [{\citenamefont {Mannakee}\ \emph {et~al.}(2016)\citenamefont
  {Mannakee}, \citenamefont {Ragsdale}, \citenamefont {Transtrum},\ and\
  \citenamefont {Gutenkunst}}]{mannakee2016sloppiness}%
  \BibitemOpen
  \bibfield  {author} {\bibinfo {author} {\bibfnamefont {B.~K.}\ \bibnamefont
  {Mannakee}}, \bibinfo {author} {\bibfnamefont {A.~P.}\ \bibnamefont
  {Ragsdale}}, \bibinfo {author} {\bibfnamefont {M.~K.}\ \bibnamefont
  {Transtrum}}, \ and\ \bibinfo {author} {\bibfnamefont {R.~N.}\ \bibnamefont
  {Gutenkunst}},\ }in\ \href@noop {} {\emph {\bibinfo {booktitle} {Uncertainty
  in Biology}}}\ (\bibinfo  {publisher} {Springer},\ \bibinfo {year} {2016})\
  pp.\ \bibinfo {pages} {271--299}\BibitemShut {NoStop}%
\bibitem [{\citenamefont {Kurniawan}\ \emph {et~al.}(2021)\citenamefont
  {Kurniawan}, \citenamefont {Petrie}, \citenamefont {Williams}, \citenamefont
  {Transtrum}, \citenamefont {Tadmor}, \citenamefont {Elliott}, \citenamefont
  {Karls},\ and\ \citenamefont {Wen}}]{kurniawan2021bayesian}%
  \BibitemOpen
  \bibfield  {author} {\bibinfo {author} {\bibfnamefont {Y.}~\bibnamefont
  {Kurniawan}}, \bibinfo {author} {\bibfnamefont {C.~L.}\ \bibnamefont
  {Petrie}}, \bibinfo {author} {\bibfnamefont {K.~J.}\ \bibnamefont
  {Williams}}, \bibinfo {author} {\bibfnamefont {M.~K.}\ \bibnamefont
  {Transtrum}}, \bibinfo {author} {\bibfnamefont {E.~B.}\ \bibnamefont
  {Tadmor}}, \bibinfo {author} {\bibfnamefont {R.~S.}\ \bibnamefont {Elliott}},
  \bibinfo {author} {\bibfnamefont {D.~S.}\ \bibnamefont {Karls}}, \ and\
  \bibinfo {author} {\bibfnamefont {M.}~\bibnamefont {Wen}},\ }\href@noop {}
  {\bibfield  {journal} {\bibinfo  {journal} {arXiv preprint arXiv:2112.10851}\
  } (\bibinfo {year} {2021})}\BibitemShut {NoStop}%
\bibitem [{\citenamefont {Marschmann}\ \emph {et~al.}(2019)\citenamefont
  {Marschmann}, \citenamefont {Pagel}, \citenamefont {K{\"u}gler},\ and\
  \citenamefont {Streck}}]{marschmann2019equifinality}%
  \BibitemOpen
  \bibfield  {author} {\bibinfo {author} {\bibfnamefont {G.~L.}\ \bibnamefont
  {Marschmann}}, \bibinfo {author} {\bibfnamefont {H.}~\bibnamefont {Pagel}},
  \bibinfo {author} {\bibfnamefont {P.}~\bibnamefont {K{\"u}gler}}, \ and\
  \bibinfo {author} {\bibfnamefont {T.}~\bibnamefont {Streck}},\ }\href@noop {}
  {\bibfield  {journal} {\bibinfo  {journal} {Environmental Modelling \&
  Software}\ }\textbf {\bibinfo {volume} {122}},\ \bibinfo {pages} {104518}
  (\bibinfo {year} {2019})}\BibitemShut {NoStop}%
\bibitem [{\citenamefont {Nik{\v{s}}i{\'c}}\ \emph {et~al.}(2017)\citenamefont
  {Nik{\v{s}}i{\'c}}, \citenamefont {Imbri{\v{s}}ak},\ and\ \citenamefont
  {Vretenar}}]{nikvsic2017sloppy}%
  \BibitemOpen
  \bibfield  {author} {\bibinfo {author} {\bibfnamefont {T.}~\bibnamefont
  {Nik{\v{s}}i{\'c}}}, \bibinfo {author} {\bibfnamefont {M.}~\bibnamefont
  {Imbri{\v{s}}ak}}, \ and\ \bibinfo {author} {\bibfnamefont {D.}~\bibnamefont
  {Vretenar}},\ }\href@noop {} {\bibfield  {journal} {\bibinfo  {journal}
  {Physical Review C}\ }\textbf {\bibinfo {volume} {95}},\ \bibinfo {pages}
  {054304} (\bibinfo {year} {2017})}\BibitemShut {NoStop}%
\bibitem [{\citenamefont {Rasband}(2021)}]{rasband2021two}%
  \BibitemOpen
  \bibfield  {author} {\bibinfo {author} {\bibfnamefont {J.}~\bibnamefont
  {Rasband}},\ }\emph {\bibinfo {title} {Two Reduced Models of Nerve
  Behavior}},\ \href@noop {} {\bibinfo {type} {Bachelor's thesis}},\ \bibinfo
  {school} {Brigham Young University} (\bibinfo {year} {2021})\BibitemShut
  {NoStop}%
\bibitem [{\citenamefont {Par{\'e}}\ \emph {et~al.}(2019)\citenamefont
  {Par{\'e}}, \citenamefont {Grimsman}, \citenamefont {Wilson}, \citenamefont
  {Transtrum},\ and\ \citenamefont {Warnick}}]{pare2019model}%
  \BibitemOpen
  \bibfield  {author} {\bibinfo {author} {\bibfnamefont {P.~E.}\ \bibnamefont
  {Par{\'e}}}, \bibinfo {author} {\bibfnamefont {D.}~\bibnamefont {Grimsman}},
  \bibinfo {author} {\bibfnamefont {A.~T.}\ \bibnamefont {Wilson}}, \bibinfo
  {author} {\bibfnamefont {M.~K.}\ \bibnamefont {Transtrum}}, \ and\ \bibinfo
  {author} {\bibfnamefont {S.}~\bibnamefont {Warnick}},\ }\href@noop {}
  {\bibfield  {journal} {\bibinfo  {journal} {IEEE Transactions on Automatic
  Control}\ }\textbf {\bibinfo {volume} {64}},\ \bibinfo {pages} {4796}
  (\bibinfo {year} {2019})}\BibitemShut {NoStop}%
\bibitem [{\citenamefont {Gerach}\ \emph {et~al.}(2019)\citenamefont {Gerach},
  \citenamefont {Wei{\ss}}, \citenamefont {D{\"o}ssel},\ and\ \citenamefont
  {Loewe}}]{gerach2019observation}%
  \BibitemOpen
  \bibfield  {author} {\bibinfo {author} {\bibfnamefont {T.}~\bibnamefont
  {Gerach}}, \bibinfo {author} {\bibfnamefont {D.}~\bibnamefont {Wei{\ss}}},
  \bibinfo {author} {\bibfnamefont {O.}~\bibnamefont {D{\"o}ssel}}, \ and\
  \bibinfo {author} {\bibfnamefont {A.}~\bibnamefont {Loewe}},\ }in\ \href@noop
  {} {\emph {\bibinfo {booktitle} {2019 Computing in Cardiology (CinC)}}}\
  (\bibinfo {organization} {IEEE},\ \bibinfo {year} {2019})\ p.~\bibinfo
  {pages} {1}\BibitemShut {NoStop}%
\bibitem [{\citenamefont {Lombardo}\ and\ \citenamefont
  {Rappel}(2017)}]{lombardo2017systematic}%
  \BibitemOpen
  \bibfield  {author} {\bibinfo {author} {\bibfnamefont {D.~M.}\ \bibnamefont
  {Lombardo}}\ and\ \bibinfo {author} {\bibfnamefont {W.-J.}\ \bibnamefont
  {Rappel}},\ }\href@noop {} {\bibfield  {journal} {\bibinfo  {journal} {Chaos:
  An Interdisciplinary Journal of Nonlinear Science}\ }\textbf {\bibinfo
  {volume} {27}},\ \bibinfo {pages} {093914} (\bibinfo {year}
  {2017})}\BibitemShut {NoStop}%
\bibitem [{\citenamefont {Par{\'e}}\ \emph {et~al.}(2015)\citenamefont
  {Par{\'e}}, \citenamefont {Wilson}, \citenamefont {Transtrum},\ and\
  \citenamefont {Warnick}}]{pare2015unified}%
  \BibitemOpen
  \bibfield  {author} {\bibinfo {author} {\bibfnamefont {P.~E.}\ \bibnamefont
  {Par{\'e}}}, \bibinfo {author} {\bibfnamefont {A.~T.}\ \bibnamefont
  {Wilson}}, \bibinfo {author} {\bibfnamefont {M.~K.}\ \bibnamefont
  {Transtrum}}, \ and\ \bibinfo {author} {\bibfnamefont {S.~C.}\ \bibnamefont
  {Warnick}},\ }in\ \href@noop {} {\emph {\bibinfo {booktitle} {2015 American
  Control Conference}}}\ (\bibinfo {organization} {IEEE},\ \bibinfo {year}
  {2015})\ pp.\ \bibinfo {pages} {1989--1994}\BibitemShut {NoStop}%
\bibitem [{\citenamefont {Lee}\ \emph {et~al.}(2003)\citenamefont {Lee},
  \citenamefont {Salic}, \citenamefont {Krüger}, \citenamefont {Heinrich},\
  and\ \citenamefont {Kirschner}}]{lee2003}%
  \BibitemOpen
  \bibfield  {author} {\bibinfo {author} {\bibfnamefont {E.}~\bibnamefont
  {Lee}}, \bibinfo {author} {\bibfnamefont {A.}~\bibnamefont {Salic}}, \bibinfo
  {author} {\bibfnamefont {R.}~\bibnamefont {Krüger}}, \bibinfo {author}
  {\bibfnamefont {R.}~\bibnamefont {Heinrich}}, \ and\ \bibinfo {author}
  {\bibfnamefont {M.~W.}\ \bibnamefont {Kirschner}},\ }\href {\doibase
  10.1371/journal.pbio.0000010} {\bibfield  {journal} {\bibinfo  {journal}
  {PLOS Biology}\ }\textbf {\bibinfo {volume} {1}},\ \bibinfo {pages} {e10}
  (\bibinfo {year} {2003})}\BibitemShut {NoStop}%
\bibitem [{\citenamefont {Ji}\ \emph {et~al.}(2019)\citenamefont {Ji},
  \citenamefont {Lu}, \citenamefont {Zamponi}, \citenamefont {Charlat},
  \citenamefont {Aversa}, \citenamefont {Yang}, \citenamefont {Sigoillot},
  \citenamefont {Zhu}, \citenamefont {Hu}, \citenamefont {Reece-Hoyes},
  \citenamefont {Russ}, \citenamefont {Michaud}, \citenamefont {Tchorz},
  \citenamefont {Jiang},\ and\ \citenamefont {Cong}}]{ji_usp7_2019}%
  \BibitemOpen
  \bibfield  {author} {\bibinfo {author} {\bibfnamefont {L.}~\bibnamefont
  {Ji}}, \bibinfo {author} {\bibfnamefont {B.}~\bibnamefont {Lu}}, \bibinfo
  {author} {\bibfnamefont {R.}~\bibnamefont {Zamponi}}, \bibinfo {author}
  {\bibfnamefont {O.}~\bibnamefont {Charlat}}, \bibinfo {author} {\bibfnamefont
  {R.}~\bibnamefont {Aversa}}, \bibinfo {author} {\bibfnamefont
  {Z.}~\bibnamefont {Yang}}, \bibinfo {author} {\bibfnamefont {F.}~\bibnamefont
  {Sigoillot}}, \bibinfo {author} {\bibfnamefont {X.}~\bibnamefont {Zhu}},
  \bibinfo {author} {\bibfnamefont {T.}~\bibnamefont {Hu}}, \bibinfo {author}
  {\bibfnamefont {J.~S.}\ \bibnamefont {Reece-Hoyes}}, \bibinfo {author}
  {\bibfnamefont {C.}~\bibnamefont {Russ}}, \bibinfo {author} {\bibfnamefont
  {G.}~\bibnamefont {Michaud}}, \bibinfo {author} {\bibfnamefont {J.~S.}\
  \bibnamefont {Tchorz}}, \bibinfo {author} {\bibfnamefont {X.}~\bibnamefont
  {Jiang}}, \ and\ \bibinfo {author} {\bibfnamefont {F.}~\bibnamefont {Cong}},\
  }\href {\doibase 10.1038/s41467-019-12143-3} {\bibfield  {journal} {\bibinfo
  {journal} {Nature Communications}\ }\textbf {\bibinfo {volume} {10}},\
  \bibinfo {pages} {4184} (\bibinfo {year} {2019})}\BibitemShut {NoStop}%
\bibitem [{\citenamefont {Gutenkunst}\ \emph {et~al.}(2007)\citenamefont
  {Gutenkunst}, \citenamefont {Waterfall}, \citenamefont {Casey}, \citenamefont
  {Brown}, \citenamefont {Myers},\ and\ \citenamefont
  {Sethna}}]{gutenkunst_universally_2007}%
  \BibitemOpen
  \bibfield  {author} {\bibinfo {author} {\bibfnamefont {R.~N.}\ \bibnamefont
  {Gutenkunst}}, \bibinfo {author} {\bibfnamefont {J.~J.}\ \bibnamefont
  {Waterfall}}, \bibinfo {author} {\bibfnamefont {F.~P.}\ \bibnamefont
  {Casey}}, \bibinfo {author} {\bibfnamefont {K.~S.}\ \bibnamefont {Brown}},
  \bibinfo {author} {\bibfnamefont {C.~R.}\ \bibnamefont {Myers}}, \ and\
  \bibinfo {author} {\bibfnamefont {J.~P.}\ \bibnamefont {Sethna}},\ }\href
  {\doibase 10.1371/journal.pcbi.0030189} {\bibfield  {journal} {\bibinfo
  {journal} {PLOS Computational Biology}\ }\textbf {\bibinfo {volume} {3}},\
  \bibinfo {pages} {e189} (\bibinfo {year} {2007})}\BibitemShut {NoStop}%
\bibitem [{\citenamefont {Transtrum}\ \emph
  {et~al.}(2011{\natexlab{b}})\citenamefont {Transtrum}, \citenamefont
  {Machta},\ and\ \citenamefont {Sethna}}]{transtrum_geometry_2011}%
  \BibitemOpen
  \bibfield  {author} {\bibinfo {author} {\bibfnamefont {M.~K.}\ \bibnamefont
  {Transtrum}}, \bibinfo {author} {\bibfnamefont {B.~B.}\ \bibnamefont
  {Machta}}, \ and\ \bibinfo {author} {\bibfnamefont {J.~P.}\ \bibnamefont
  {Sethna}},\ }\href {\doibase 10.1103/PhysRevE.83.036701} {\bibfield
  {journal} {\bibinfo  {journal} {Physical Review E}\ }\textbf {\bibinfo
  {volume} {83}},\ \bibinfo {pages} {036701} (\bibinfo {year}
  {2011}{\natexlab{b}})}\BibitemShut {NoStop}%
\bibitem [{\citenamefont {Pourquie}(2003)}]{Pourquie2003}%
  \BibitemOpen
  \bibfield  {author} {\bibinfo {author} {\bibfnamefont {O.}~\bibnamefont
  {Pourquie}},\ }\href {\doibase 10.1126/science.1085887} {\bibfield  {journal}
  {\bibinfo  {journal} {Science}\ }\textbf {\bibinfo {volume} {301}},\ \bibinfo
  {pages} {328} (\bibinfo {year} {2003})}\BibitemShut {NoStop}%
\bibitem [{\citenamefont {Caruana}(1997)}]{Caruana1997}%
  \BibitemOpen
  \bibfield  {author} {\bibinfo {author} {\bibfnamefont {R.}~\bibnamefont
  {Caruana}},\ }\href {\doibase 10.1023/A:1007379606734} {\bibfield  {journal}
  {\bibinfo  {journal} {Machine Learning}\ }\textbf {\bibinfo {volume} {28}},\
  \bibinfo {pages} {41} (\bibinfo {year} {1997})}\BibitemShut {NoStop}%
\bibitem [{\citenamefont {Rumelhart}\ and\ \citenamefont
  {Abrahamson}(1973)}]{rumelhart1973model}%
  \BibitemOpen
  \bibfield  {author} {\bibinfo {author} {\bibfnamefont {D.~E.}\ \bibnamefont
  {Rumelhart}}\ and\ \bibinfo {author} {\bibfnamefont {A.~A.}\ \bibnamefont
  {Abrahamson}},\ }\href {\doibase 10.1016/0010-0285(73)90023-6} {\bibfield
  {journal} {\bibinfo  {journal} {Cognitive Psychology}\ }\textbf {\bibinfo
  {volume} {5}},\ \bibinfo {pages} {1} (\bibinfo {year} {1973})}\BibitemShut
  {NoStop}%
\bibitem [{\citenamefont {Peterson}\ \emph {et~al.}(2020)\citenamefont
  {Peterson}, \citenamefont {Chen},\ and\ \citenamefont
  {Griffiths}}]{peterson2020parallelograms}%
  \BibitemOpen
  \bibfield  {author} {\bibinfo {author} {\bibfnamefont {J.~C.}\ \bibnamefont
  {Peterson}}, \bibinfo {author} {\bibfnamefont {D.}~\bibnamefont {Chen}}, \
  and\ \bibinfo {author} {\bibfnamefont {T.~L.}\ \bibnamefont {Griffiths}},\
  }\href {\doibase 10.1016/j.cognition.2020.104440} {\bibfield  {journal}
  {\bibinfo  {journal} {Cognition}\ }\textbf {\bibinfo {volume} {205}},\
  \bibinfo {pages} {104440} (\bibinfo {year} {2020})}\BibitemShut {NoStop}%
\bibitem [{\citenamefont {Musto}(2010)}]{musto2010enhanced}%
  \BibitemOpen
  \bibfield  {author} {\bibinfo {author} {\bibfnamefont {C.}~\bibnamefont
  {Musto}},\ }in\ \href@noop {} {\emph {\bibinfo {booktitle} {Proceedings of
  the fourth ACM conference on Recommender systems}}}\ (\bibinfo {year}
  {2010})\ pp.\ \bibinfo {pages} {361--364}\BibitemShut {NoStop}%
\bibitem [{\citenamefont {Reid}\ and\ \citenamefont
  {Katz}(2018)}]{reid2018vector}%
  \BibitemOpen
  \bibfield  {author} {\bibinfo {author} {\bibfnamefont {J.~N.}\ \bibnamefont
  {Reid}}\ and\ \bibinfo {author} {\bibfnamefont {A.~N.}\ \bibnamefont
  {Katz}},\ }\href {\doibase 10.1080/10926488.2018.1549840} {\bibfield
  {journal} {\bibinfo  {journal} {Metaphor and Symbol}\ }\textbf {\bibinfo
  {volume} {33}},\ \bibinfo {pages} {280} (\bibinfo {year} {2018})}\BibitemShut
  {NoStop}%
\bibitem [{\citenamefont {Gentner}(1983)}]{Gentner1983}%
  \BibitemOpen
  \bibfield  {author} {\bibinfo {author} {\bibfnamefont {D.}~\bibnamefont
  {Gentner}},\ }\href {\doibase 10.1016/S0364-0213(83)80009-3} {\bibfield
  {journal} {\bibinfo  {journal} {Cognitive Science}\ }\textbf {\bibinfo
  {volume} {7}},\ \bibinfo {pages} {155} (\bibinfo {year} {1983})}\BibitemShut
  {NoStop}%
\end{thebibliography}%


\begin{thebibliography}{6}%
\makeatletter
\providecommand \@ifxundefined [1]{%
 \@ifx{#1\undefined}
}%
\providecommand \@ifnum [1]{%
 \ifnum #1\expandafter \@firstoftwo
 \else \expandafter \@secondoftwo
 \fi
}%
\providecommand \@ifx [1]{%
 \ifx #1\expandafter \@firstoftwo
 \else \expandafter \@secondoftwo
 \fi
}%
\providecommand \natexlab [1]{#1}%
\providecommand \enquote  [1]{``#1''}%
\providecommand \bibnamefont  [1]{#1}%
\providecommand \bibfnamefont [1]{#1}%
\providecommand \citenamefont [1]{#1}%
\providecommand \href@noop [0]{\@secondoftwo}%
\providecommand \href [0]{\begingroup \@sanitize@url \@href}%
\providecommand \@href[1]{\@@startlink{#1}\@@href}%
\providecommand \@@href[1]{\endgroup#1\@@endlink}%
\providecommand \@sanitize@url [0]{\catcode `\\12\catcode `\$12\catcode
  `\&12\catcode `\#12\catcode `\^12\catcode `\_12\catcode `\%12\relax}%
\providecommand \@@startlink[1]{}%
\providecommand \@@endlink[0]{}%
\providecommand \url  [0]{\begingroup\@sanitize@url \@url }%
\providecommand \@url [1]{\endgroup\@href {#1}{\urlprefix }}%
\providecommand \urlprefix  [0]{URL }%
\providecommand \Eprint [0]{\href }%
\providecommand \doibase [0]{http://dx.doi.org/}%
\providecommand \selectlanguage [0]{\@gobble}%
\providecommand \bibinfo  [0]{\@secondoftwo}%
\providecommand \bibfield  [0]{\@secondoftwo}%
\providecommand \translation [1]{[#1]}%
\providecommand \BibitemOpen [0]{}%
\providecommand \bibitemStop [0]{}%
\providecommand \bibitemNoStop [0]{.\EOS\space}%
\providecommand \EOS [0]{\spacefactor3000\relax}%
\providecommand \BibitemShut  [1]{\csname bibitem#1\endcsname}%
\let\auto@bib@innerbib\@empty
\bibitem [{\citenamefont {Nusse}\ and\ \citenamefont
  {Clevers}(2017)}]{Nusse2017}%
  \BibitemOpen
  \bibfield  {author} {\bibinfo {author} {\bibfnamefont {R.}~\bibnamefont
  {Nusse}}\ and\ \bibinfo {author} {\bibfnamefont {H.}~\bibnamefont
  {Clevers}},\ }\href {\doibase 10.1016/j.cell.2017.05.016} {\bibfield
  {journal} {\bibinfo  {journal} {Cell}\ }\textbf {\bibinfo {volume} {169}},\
  \bibinfo {pages} {985} (\bibinfo {year} {2017})}\BibitemShut {NoStop}%
\bibitem [{\citenamefont {Logan}\ and\ \citenamefont
  {Nusse}(2004)}]{logan_wnt_2004}%
  \BibitemOpen
  \bibfield  {author} {\bibinfo {author} {\bibfnamefont {C.~Y.}\ \bibnamefont
  {Logan}}\ and\ \bibinfo {author} {\bibfnamefont {R.}~\bibnamefont {Nusse}},\
  }\href {\doibase 10.1146/annurev.cellbio.20.010403.113126} {\bibfield
  {journal} {\bibinfo  {journal} {Annual Review of Cell and Developmental
  Biology}\ }\textbf {\bibinfo {volume} {20}},\ \bibinfo {pages} {781}
  (\bibinfo {year} {2004})}\BibitemShut {NoStop}%
\bibitem [{\citenamefont {Ding}\ \emph {et~al.}(2014)\citenamefont {Ding},
  \citenamefont {Su}, \citenamefont {Tang}, \citenamefont {Zhang},
  \citenamefont {Chen}, \citenamefont {Zhu}, \citenamefont {Liang},
  \citenamefont {Wei}, \citenamefont {Guo}, \citenamefont {Liu}, \citenamefont
  {Chen},\ and\ \citenamefont {Wu}}]{ding_enrichment_2014}%
  \BibitemOpen
  \bibfield  {author} {\bibinfo {author} {\bibfnamefont {Y.}~\bibnamefont
  {Ding}}, \bibinfo {author} {\bibfnamefont {S.}~\bibnamefont {Su}}, \bibinfo
  {author} {\bibfnamefont {W.}~\bibnamefont {Tang}}, \bibinfo {author}
  {\bibfnamefont {X.}~\bibnamefont {Zhang}}, \bibinfo {author} {\bibfnamefont
  {S.}~\bibnamefont {Chen}}, \bibinfo {author} {\bibfnamefont {G.}~\bibnamefont
  {Zhu}}, \bibinfo {author} {\bibfnamefont {J.}~\bibnamefont {Liang}}, \bibinfo
  {author} {\bibfnamefont {W.}~\bibnamefont {Wei}}, \bibinfo {author}
  {\bibfnamefont {Y.}~\bibnamefont {Guo}}, \bibinfo {author} {\bibfnamefont
  {L.}~\bibnamefont {Liu}}, \bibinfo {author} {\bibfnamefont {Y.-G.}\
  \bibnamefont {Chen}}, \ and\ \bibinfo {author} {\bibfnamefont
  {W.}~\bibnamefont {Wu}},\ }\href {\doibase 10.1242/jcs.146977} {\bibfield
  {journal} {\bibinfo  {journal} {Journal of Cell Science}\ }\textbf {\bibinfo
  {volume} {127}},\ \bibinfo {pages} {4833} (\bibinfo {year}
  {2014})}\BibitemShut {NoStop}%
\bibitem [{\citenamefont {Lee}\ \emph {et~al.}(2003)\citenamefont {Lee},
  \citenamefont {Salic}, \citenamefont {Krüger}, \citenamefont {Heinrich},\
  and\ \citenamefont {Kirschner}}]{lee2003}%
  \BibitemOpen
  \bibfield  {author} {\bibinfo {author} {\bibfnamefont {E.}~\bibnamefont
  {Lee}}, \bibinfo {author} {\bibfnamefont {A.}~\bibnamefont {Salic}}, \bibinfo
  {author} {\bibfnamefont {R.}~\bibnamefont {Krüger}}, \bibinfo {author}
  {\bibfnamefont {R.}~\bibnamefont {Heinrich}}, \ and\ \bibinfo {author}
  {\bibfnamefont {M.~W.}\ \bibnamefont {Kirschner}},\ }\href {\doibase
  10.1371/journal.pbio.0000010} {\bibfield  {journal} {\bibinfo  {journal}
  {PLOS Biology}\ }\textbf {\bibinfo {volume} {1}},\ \bibinfo {pages} {e10}
  (\bibinfo {year} {2003})}\BibitemShut {NoStop}%
\bibitem [{\citenamefont {Goentoro}\ and\ \citenamefont
  {Kirschner}(2009)}]{goentoro2009}%
  \BibitemOpen
  \bibfield  {author} {\bibinfo {author} {\bibfnamefont {L.}~\bibnamefont
  {Goentoro}}\ and\ \bibinfo {author} {\bibfnamefont {M.~W.}\ \bibnamefont
  {Kirschner}},\ }\href {\doibase 10.1016/j.molcel.2009.11.017} {\bibfield
  {journal} {\bibinfo  {journal} {Molecular Cell}\ }\textbf {\bibinfo {volume}
  {36}},\ \bibinfo {pages} {872} (\bibinfo {year} {2009})}\BibitemShut
  {NoStop}%
\bibitem [{\citenamefont {Jensen}\ \emph {et~al.}(2010)\citenamefont {Jensen},
  \citenamefont {Pedersen}, \citenamefont {Krishna},\ and\ \citenamefont
  {Jensen}}]{jensen2010}%
  \BibitemOpen
  \bibfield  {author} {\bibinfo {author} {\bibfnamefont {P.~B.}\ \bibnamefont
  {Jensen}}, \bibinfo {author} {\bibfnamefont {L.}~\bibnamefont {Pedersen}},
  \bibinfo {author} {\bibfnamefont {S.}~\bibnamefont {Krishna}}, \ and\
  \bibinfo {author} {\bibfnamefont {M.~H.}\ \bibnamefont {Jensen}},\ }\href
  {\doibase 10.1016/j.bpj.2009.11.039} {\bibfield  {journal} {\bibinfo
  {journal} {Biophysical Journal}\ }\textbf {\bibinfo {volume} {98}},\ \bibinfo
  {pages} {943} (\bibinfo {year} {2010})}\BibitemShut {NoStop}%
\end{thebibliography}%

\end{document}


{\center{\section*{\Large{Supplementary Material}}}}

\section{Spin model with ferromagnetic coupling}
\begin{figure}[h!]
    \centering
    \includegraphics[width=3.5in]{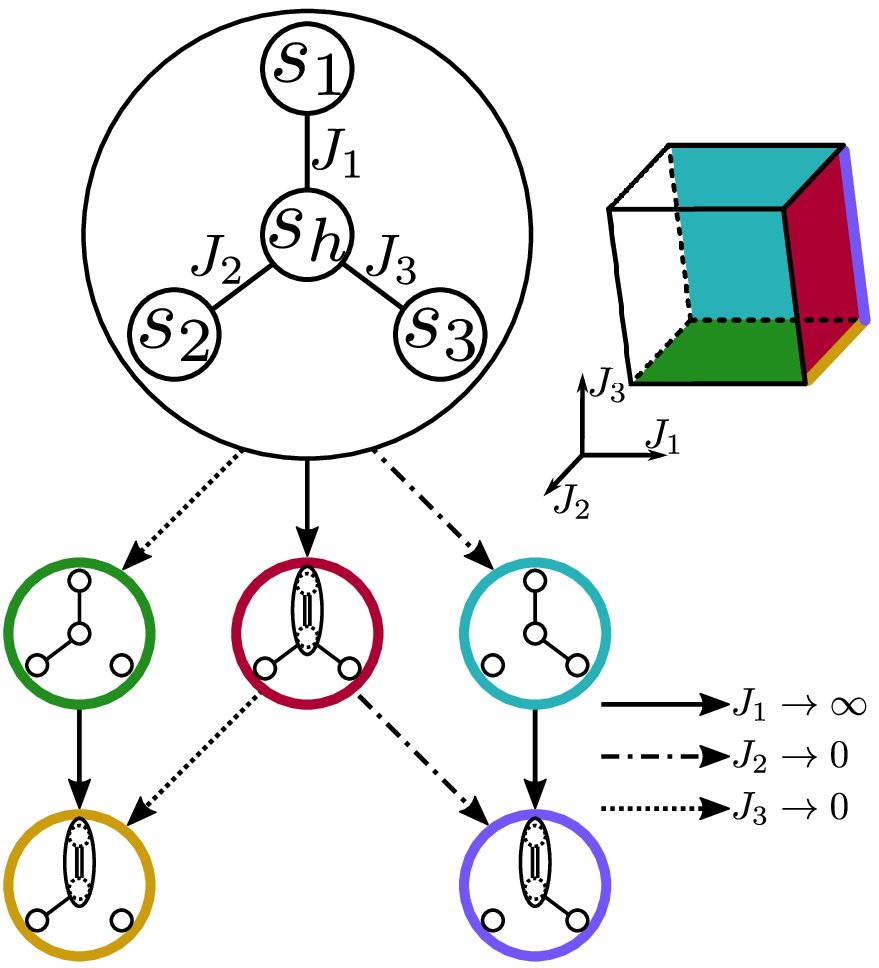}
    \caption{
    Hasse diagram showing the POSet of reduced models, including the supremum, for spin model with ferromagnetic coupling on a network.
    Observables for the left and right flags of Hasse diagram are $P(s_1,s_2)$ and $P(s_1,s_3)$, respectively.
    The model manifold is diffeomorphic to a cube, which is illustrated in the top right corner.
    The colored faces and edges represent reduced models, while the red face represents the supremum.
    }
    \label{fig:spin}
\end{figure}

Here, we demonstrate the supremum principle with another simple model, a spin model with ferromagnetic coupling as illustrated in the top node of the Hasse diagram in Fig.~\ref{fig:spin}.
Spin variables can take on values $s \in {-1, 0. 1}$.
The model manifold shares the same topology as a cube and is represented in the top right of Fig.~\ref{fig:spin}.
The energy function corresponding to the top node in the Hasse diagram of Fig.~\ref{fig:spin} is
\begin{equation}
    H = -J_1s_1s_h - J_2s_2s_h - J_3s_3s_h.
\end{equation}
The full probability distribution, marginalized over the hidden spin, is given by
\begin{align}
    P(s_1,s_2,s_3)  &= \frac{1}{Z} \sum\limits_{s_h=-1,0,1} e^{J_1s_1s_h + J_2s_2s_h + J_3s_3s_h} \\
                    &= \frac{1}{Z} \left(1+2\cosh\left(J_1s_1+J_2s_2+J_3s_3\right)\right),
    \label{eq:p123}
\end{align}
with the partition function being given by
\begin{equation}
    Z = 29+4\cosh(J_3)+4\cosh(J_1)\left(1+2\cosh(J_2)\right)\left(1+2\cosh(J_3)\right)+\cosh(J_2)\left(4+8\cosh(J_3)\right).
\end{equation}
We consider the case where $s_h$ couples strongly to $s_1$ but weakly to $s_2$ and $s_3$, and experiments can only measure the correlations $P(s_1,s_2)$ and $P(s_1,s_3)$.
The probability for only two spins can be found by marginalizing over one spin and is given by
\begin{equation}
    P(s_i,s_j) = \sum\limits_{s_k=-1,0,1} P(s_i,s_j,s_k),
\end{equation}
where $i$ and $j$ can be any discrete pair of $1, 2, 3$ and $k$ is the remaining spin.

\subsection{Possible MBAM reductions}
For ferromagnetic spin systems there are two types of possible reductions that MBAM can find; taking a single $J_i$ to zero or infinity.
To illustrate each type of reduction, consider taking $J_1$ to each of the extreme values for the observable $P(s_1,s_2,s_3)$ in Eq.~\ref{eq:p123}.
Taking $J_0$ to zero is trivial, the two correlated spins are decoupled resulting in the limit
\begin{equation}
    \lim\limits_{J_1\rightarrow 0} P(s_1,s_2,s_3) = \frac{\left(1+2\cosh\left(J_2s_2+J_3s_3\right)\right)}{33+12\cosh(J_2)+12\cosh(J_3)+24\cosh(J_2)\cosh(J_3)}.
\end{equation}
In this case the spins $s_1$, $s_2$, and $s_3$ can all take on the values $-1$, $0$, and $1$.

The limit $J_1\rightarrow\infty$ is more subtle.
When taking this limit, it is useful to set $s_1$ to each of it's possible values $-1$, $0$, and $1$ before taking the limit.
The reduced model for each probability is given by
\begin{equation}
    \lim\limits_{J_1\rightarrow 0} P(s_1=\{-1,0,1\},s_2,s_3) = \frac{\left\{e^{-J_2s_2-J_3s_3},0,e^{J_2s_2+J_3s_3}\right\}}{2+4\cosh(J_2)+4\cosh(J_3)+8\cosh(J_2)\cosh(J_3)}.
\end{equation}
The configurations with $s_1=0$ all have a probability of zero.
Thus, this is equivalent to the model 
\begin{equation}
    \lim\limits_{J_1\rightarrow 0} P(s_1,s_2,s_3) = \frac{e^{J_2s_2s_1+J_3s_3s_1}}{2+4\cosh(J_2)+4\cosh(J_3)+8\cosh(J_2)\cosh(J_3)},
\end{equation}
where $s_2$ and $s_3$ can be either $-1$, $0$, or $1$, but $s_1$ is restricted to only be $-1$ or $1$.
In this way, reducing the model not only decreases the number of parameters, but also reduces the number of configurations with non-zero probability.

\subsection{Reduced model based on $P(s_1,s_2)$ data}
We first consider the case where only $s_1$ and $s_2$ can be measured.
Observing $P(s_1,s_2)$ renders $J_3$ practically unidentifiable, represented by the limit $J_3 \rightarrow 0$ and the green face of the model manifold.
Further, the strong correlation between $s_1$ and $s_h$ is represented by the limit $J_1 \rightarrow \infty$ and the gold edge of the model manifold.
Applying these limits to $P(s_1,s_2)$ yields the reduced model
\begin{align}
    \lim_{\substack{J_3\rightarrow 0 \\ J_1\rightarrow\infty}} P(s_1,s_2) = \frac{e^{J_2s_2s_1}}{2+4\cosh(J_2)}.
\end{align}
This is the gold model in the Hasse diagram of Fig.~\ref{fig:spin} with the energy function
\begin{equation}
    H_\text{gold} = -J_2s_2s_1,
\end{equation}
with $s_1$ confined to be only -1, 1, as all configurations with $s_1=0$ are zero in this limit.

\subsection{Reduced model based on $P(s_1,s_3)$ data}
If we consider the case where only $s_1$ and $s_3$ can be measured.
By the similar arguments made in the previous section, the relevant limits were $J_2\rightarrow 0$, $J_1\rightarrow\infty$.
Applying these limits to $P(s_1,s_3)$ yields the reduced model
\begin{align}
    \lim_{\substack{J_2\rightarrow 0 \\ J_1\rightarrow\infty}} P(s_1,s_3) = \frac{e^{J_3s_3s_1}}{2+4\cosh(J_3)}.
\end{align}
This is the purple model in the Hasse diagram of Fig.~\ref{fig:spin} with the energy function
\begin{equation}
    H_\text{purple} = -J_3s_3s_1.
\end{equation}
Again, $s_1$ is confined to be only -1, 1.

\subsection{Building the supremum}
We construct the supremum by applying the overlapping parameter reductions, only $J_1 \rightarrow \infty$ in this simple case, to the original full model.
Applying this limit to the full probability function $P(s_1,s_2,s_3)$ yields the supremum:
\begin{align}
    P_\text{sup}(s_1,s_2,s_3)    &= \lim\limits_{J_1\rightarrow\infty} P(s_1,s_2,s_3) \\
                            &= \frac{e^{J_2s_2s_1+J_3s_3s_1}}{Z},
\end{align}
where the partition function is given by
\begin{equation}
    Z = 1+4\cosh(J_2)+4\cosh(J_3)+8\cosh(J_2)\cosh(J_3).
\end{equation}
In this limit $s_1$ is again restricted to be -1 and 1.
This is the supremal model shown in red in Fig.~\ref{fig:spin}, having the energy function
\begin{equation}
    H_\text{sup} = -J_2s_2s_1 - J_3s_3s_1,
\end{equation}
with $s_1={-1,1}$ and $s_2,s_3={-1,0,1}$.
The red face is the lowest dimensional face that contains the gold and purple edges.
The supremum contains the information to predict, not only $P(s_1,s_2)$ and $P(s_1,s_3)$, but the experimentally inaccessible distributions, $P(s_2,s_3)$ and $P(s_1,s_2,s_3)$.

\section{\label{apx:Wnt}Wnt signaling pathway}

The Wnt signaling pathway is the dominant mechanism for initiating cell division in almost all animals \cite{Nusse2017}.
This relaying of a local signal to the nucleus is crucial to normal embryonic development, stem-cell activation, and cancer tumorigenesis, and thus is one of the best-studied in all of biology.
The canonical Wnt pathway is a many-step process, summarized in Fig.~2a.
First, one of several extracellular Wnt molecules (such as the eponymous Wingless-or-Int-1 proteins, or one of their many analogues) interacts with two intermembrane proteins: Frizzled and LRP, forming a complex (L).
Inside the cell, L binds to Axin (A), thereby removing it from the APC-GSK3-Axin destruction complex (DC) which normally degrades $\beta$-catenin ($\beta$).
In the absence of DC, $\beta$-catenin now accumulates and interacts with DNA-associated proteins such as TCF to promote cell division \cite{logan_wnt_2004, ding_enrichment_2014}
This ``accumulation phenomenon'' is well-documented in the literature \cite{lee2003, goentoro2009} and illustrated in Fig.~2b.

A mass balance model of even this simple outline of the pathway contains over a dozen parameters, obscuring the relationship among output behaviors of $\beta$-catenin for different mechanisms.
The problem is compounded by the fact that $\beta$-catenin is a transcription factor for many different genes depending on the state of the cell.
For example, during somitogenesis $\beta$-catenin activates Axin2 (a homolog of Axin) leading to a negative feedback loop, driving a limit cycle ``oscillation phenomenon'' that acts as a segmentation clock \cite{jensen2010} (see Fig.~2b).

Various models for the Wnt signaling pathway were discussed in the main text.
We will give the equations for each of the most relevant models, i.e., those in Fig.~3a.
In the equations, the dynamic equations will be
\begin{equation}
\begin{split}
    y_1 &= \text{Destruction Complex}[\text{GSK-Axin2-}\beta] \\
    y_2 &= \text{[GSK-Axin2]} \\
    y_3 &= [\beta-\text{catenin}] \\
    y_4 &= \text{[GSK]} \\
    y_5 &= \text{[Axin2]} \\
    y_6 &= [\text{Axin2}_{\text{mRNA}}\text{]} \\
    y_7 &= \text{[Axin2-LRP]} \\
    y_8 &= \text{[LRP]},
\end{split}
\end{equation}
where $\beta$-catenin is the observed variable in each case, and common renormalized quantities are
\begin{equation}
\begin{split}
    \widetilde{y}_6 &= c_{tlA} y_6 \\
    k_{GA} &= c_{bGA}/c_{fGA}.
\end{split}
\end{equation}
Additionally, the input function for the Wnt and USP7 are given by
\begin{equation}
\begin{split}
    f_W(t) &= 70.0\times0.0398942 e^{\frac{-(t-200)^2}{2\times10^2}} \\
    f_U(t) &= 
    \begin{cases}
    15.0 e^{\frac{-(t-1050)^2}{2\times10^2}},& \text{Interrupted} \\
    1,              & \text{Accumulation},
\end{cases}
\end{split}
\end{equation}
and time derivatives of the variables are $\dot{y} = dy/dt$.

\subsection*{A14: Accumulation Model, $N=$14}
\begin{equation}
\begin{split}
    \dot{y}_1 &= c_{fC} y_3 y_2 - c_{bC} y_1 - \alpha y_1 \\
    \dot{y}_2 &= c_{fGA} y_4 y_5 - c_{bGA} y_2 - c_{fC} y_3 y_2 + c_{bC} y_1 + \alpha y_1 \\
    \dot{y}_3 &= S - c_{fC} y_3 y_2 + c_{bC} y_1 \\
    \dot{y}_4 &= -c_{fGA} y_4 y_5 + c_{bGA} y_2 \\
    \dot{y}_5 &= -c_{fGA} y_4 y_5 + c_{bGA} y_2 + c_{tlA} y_6 - c_{fAL} y_5 y_8 + c_{bAL} y_7 \\
    \dot{y}_6 &= -1/\tau_{Am} y_6 + SA \\
    \dot{y}_7 &= c_{fAL} y_5 y_8 - c_{bAL} y_7 - \nu y_7 \\
    \dot{y}_8 &= -c_{fAL} y_5 y_8 + c_{bAL} y_7 + \nu y_7 + f_W(t) \\
    \text{Conserved Quantities:}& \\
    G_{tot} & = y_1 + y_2 + y_4 \\
    L_{tot} &= y_7 + y_8
\end{split}
\label{eq:A14}
\end{equation}

\subsection*{A9: Accumulation Model, $N=$9}
\begin{equation}
\begin{split}
    \dot{y}_2 &= \frac{y_4\left(\widetilde{y}_6 - c_{fAL} y_5 y_8\right)}{k_{GA} + y_4 + y_5} \\
    \dot{y}_3 &= S - c_{fC} y_3 y_2 \\
    \dot{y}_4 &= -\frac{y_4\left(\widetilde{y}_6 - c_{fAL} y_5 y_8\right)}{k_{GA} + y_4 + y_5} \\
    \dot{y}_5 &= -\frac{y_4\left(\widetilde{y}_6 - c_{fAL} y_5 y_8\right)}{k_{GA} + y_4 + y_5} + \widetilde{y}_6 - c_{fAL} y_5 y_8 \\
    \dot{\widetilde{y}}_6 &= -1/\tau_{Am} \widetilde{y}_6 + \left[SA\cdot c_{tlA}\right] \\
    \dot{y}_7 &= c_{fAL} y_5 y_8 - \nu y_7 \\
    \dot{y}_8 &= -c_{fAL} y_5 y_8 + \nu y_7 + f_W(t) \\
    \text{IC conditions:}& \\
    G_{tot} & = y_2 + y_4 \\
    L_{tot} &= y_7 + y_8
\end{split}
\label{eq:A9}
\end{equation}

\subsection*{A3: Accumulation Model, $N=$3}
\begin{equation}
\begin{split}
    \dot{y}_3 &= S - \theta_1 \frac{y_3}{y_8} \\
    \dot{y}_8 &= f_W(t) \\
    \text{Conserved Quantities:}& \\
    y_8 &= \theta_2 \\
    \text{Renormalized parameters:}& \\
    \theta_1 &= \frac{c_{fC}\cdot c_{fGA}\cdot SA\cdot c_{tlA}\cdot \tau_{Am}}{G_{tot}\cdot c_{fAL}\cdot c_{bGA}} \\
    \theta_2 &= L_{tot} - \frac{SA\cdot c_{tlA}\cdot \tau_{Am}}{\nu}
\end{split}
\label{eq:A3}
\end{equation}

\subsection*{O14: Oscillation Model, $N=$14}
\begin{equation}
\begin{split}
    \dot{y}_1 &= c_{fC} y_3 y_2 - c_{bC} y_1 - \alpha y_1 \\
    \dot{y}_2 &= c_{fGA} y_4 y_5 - c_{bGA} y_2 - c_{fC} y_3 y_2 + c_{bC} y_1 + \alpha y_1 \\
    \dot{y}_3 &= S - c_{fC} y_3 y_2 + c_{bC} y_1 \\
    \dot{y}_4 &= -c_{fGA} y_4 y_5 + c_{bGA} y_2 \\
    \dot{y}_5 &= -c_{fGA} y_4 y_5 + c_{bGA} y_2 + c_{tlA} y_6 - c_{fAL} y_5 y_8 + c_{bAL} y_7 \\
    \dot{y}_6 &= -1/\tau_{Am} y_6 + c_{tsA} y_3^2 \\
    \dot{y}_7 &= c_{fAL} y_5 y_8 - c_{bAL} y_7 - \nu y_7 \\
    \dot{y}_8 &= -c_{fAL} y_5 y_8 + c_{bAL} y_7 + \nu y_7 + f_W(t) \\
    \text{Conserved Quantities:}& \\
    G_{tot} & = y_1 + y_2 + y_4 \\
    L_{tot} &= y_7 + y_8
\end{split}
\label{eq:O14}
\end{equation}

\subsection*{O9: Oscillation Model, $N=$9}
\begin{equation}
\begin{split}
    \dot{y}_2 &= \frac{y_4\left(\widetilde{y}_6 - c_{fAL} y_5 y_8\right)}{k_{GA} + y_4 + y_5} \\
    \dot{y}_3 &= S - c_{fC} y_3 y_2 \\
    \dot{y}_4 &= -\frac{y_4\left(\widetilde{y}_6 - c_{fAL} y_5 y_8\right)}{k_{GA} + y_4 + y_5} \\
    \dot{y}_5 &= -\frac{y_4\left(\widetilde{y}_6 - c_{fAL} y_5 y_8\right)}{k_{GA} + y_4 + y_5} + \widetilde{y}_6 - c_{fAL} y_5 y_8 \\
    \dot{\widetilde{y}}_6 &= -1/\tau_{Am} \widetilde{y}_6 + \left[c_{tsA}\cdot c_{tlA}\right] y_3^2 \\
    \dot{y}_7 &= c_{fAL} y_5 y_8 - \nu y_7 \\
    \dot{y}_8 &= -c_{fAL} y_5 y_8 + \nu y_7 + f_W(t) \\
    \text{Conserved Quantities:}& \\
    G_{tot} & = y_2 + y_4 \\
    L_{tot} &= y_7 + y_8
\end{split}
\label{eq:O9}
\end{equation}

\subsection*{C15: Combined Model, $N=$15}
\begin{equation}
\begin{split}
    \dot{y}_1 &= c_{fC} y_3 y_2 - c_{bC} y_1 - \alpha y_1 \\
    \dot{y}_2 &= c_{fGA} y_4 y_5 - c_{bGA} y_2 - c_{fC} y_3 y_2 + c_{bC} y_1 + \alpha y_1 \\
    \dot{y}_3 &= S - c_{fC} y_3 y_2 + c_{bC} y_1 \\
    \dot{y}_4 &= -c_{fGA} y_4 y_5 + c_{bGA} y_2 \\
    \dot{y}_5 &= -c_{fGA} y_4 y_5 + c_{bGA} y_2 + c_{tlA} y_6 - c_{fAL} y_5 y_8 + c_{bAL} y_7 \\
    \dot{y}_6 &= -1/\tau_{Am} y_6 + c_{tsA} y_3^2 + SA f_U(t)\\
    \dot{y}_7 &= c_{fAL} y_5 y_8 - c_{bAL} y_7 - \nu y_7 \\
    \dot{y}_8 &= -c_{fAL} y_5 y_8 + c_{bAL} y_7 + \nu y_7 + f_W(t) \\
    \text{Conserved Quantities:}& \\
    G_{tot} & = y_1 + y_2 + y_4 \\
    L_{tot} &= y_7 + y_8
\end{split}
\label{eq:C15}
\end{equation}

\subsection*{C10: Supremum Model, $N=$10}
\begin{equation}
\begin{split}
    \dot{y}_2 &= \frac{y_4\left(\widetilde{y}_6 - c_{fAL} y_5 y_8\right)}{k_{GA} + y_4 + y_5} \\
    \dot{y}_3 &= S - c_{fC} y_3 y_2 \\
    \dot{y}_4 &= -\frac{y_4\left(\widetilde{y}_6 - c_{fAL} y_5 y_8\right)}{k_{GA} + y_4 + y_5} \\
    \dot{y}_5 &= -\frac{y_4\left(\widetilde{y}_6 - c_{fAL} y_5 y_8\right)}{k_{GA} + y_4 + y_5} + \widetilde{y}_6 - c_{fAL} y_5 y_8 \\
    \dot{\widetilde{y}}_6 &= -1/\tau_{Am} \widetilde{y}_6 + \left[c_{tsA}\cdot c_{tlA}\right] y_3^2 + \left[SA\cdot c_{tlA}\right] f_U(t) \\
    \dot{y}_7 &= c_{fAL} y_5 y_8 - \nu y_7 \\
    \dot{y}_8 &= -c_{fAL} y_5 y_8 + \nu y_7 + f_W(t) \\
    \text{Conserved Quantities:}& \\
    G_{tot} & = y_2 + y_4 \\
    L_{tot} &= y_7 + y_8
\end{split}
\label{eq:C10}
\end{equation}

Tables~\ref{tab:redgreenlimits}-\ref{tab:blacklimits} show the limits for the various sequences of reductions, illustrated by the colored arrows in Fig.~3a.
As mentioned in the text, given the correct parameterization, each limit can be described by taking a single parameter to zero.
These reparameterized parameters are shown in Table~\ref{tab:bluelimits} for the common limits that are used in the construction of the supremum, indicated by the blue arrow in Fig.~3a.
Table~\ref{tab:rep_limits} shows the original and reparameterized limits side by side.

\begin{table}[H]
\centering
\begin{tabular}{ |c|c| } 
    \hline
    Red & Green \\
    \hline
    $c_{tSA} \rightarrow 0$ & $SA \rightarrow 0$ \\
    \hline
\end{tabular}
\caption{Sequence of limits represented by the red and green arrows in Fig. 3.}
\label{tab:redgreenlimits}
\end{table}

\begin{table}[H]
\centering
\begin{tabular}{ |c|c| }
    \hline
    Blue & Reparameterized Blue \\
    \hline
    $c_{bC} \rightarrow 0$ & 
    $c_{bC} \rightarrow 0$ \\
    
    \hline
    $\alpha \rightarrow \infty$ &
    $\frac{1}{\alpha} \rightarrow 0$ \\
    
    \hline
    \begin{tabular}{ c c }
    $c_{tlA} \rightarrow \infty, SA \rightarrow 0$ & (Accumulation) \\
    $c_{tlA} \rightarrow \infty, c_{tsA} \rightarrow 0$ & (Oscillation) \\
    $c_{tlA} \rightarrow \infty,  SA \rightarrow 0, c_{tsA} \rightarrow 0$ & (Interrupted)
    \end{tabular} &
    $\frac{1}{c_{tlA}} \rightarrow 0$ \\
    
    \hline
    $c_{bAL} \rightarrow 0$ &
    $c_{bAL} \rightarrow 0$ \\
    
    \hline
    \begin{tabular}{ c }
        $c_{bGA} \rightarrow \infty$ \\
        $c_{fGA} \rightarrow \infty$
    \end{tabular} &
    $\frac{1}{c_{fGA}} \rightarrow 0$ \\
    \hline
\end{tabular}
\caption{Sequence of limits represented by the blue arrow in Fig. 3. The second column shows the limits with renormalized parameters such that each limit is given by a single parameter going to zero. }
\label{tab:bluelimits}
\end{table}

\begin{table}[H]
\centering
\begin{tabular}{ |c| } 
    \hline
    Orange \\
    \hline
    \begin{tabular}{ c }
        $SA\cdot c_{tlA} \rightarrow 0$ \\
        $\tau_{Am} \rightarrow \infty$
    \end{tabular} \\
    \hline
    
    \begin{tabular}{ c }
        $\frac{c_{bGA}}{c_{fGA}} \rightarrow \infty$ \\
        $G_{tot} \rightarrow \infty$
    \end{tabular} \\
    \hline
    
    \begin{tabular}{ c }
        $\frac{G_{tot}\cdot c_{fGA}}{c_{bGA}} \rightarrow \infty$ \\
        $c_{fAL} \rightarrow \infty$
    \end{tabular} \\
    \hline
    
    \begin{tabular}{ c }
        $SA\cdot c_{tlA}\cdot \tau_{Am} \rightarrow \infty$ \\
        $\frac{G_{tot}\cdot c_{fGA}}{c_{bGA}\cdot c_{fAL}} \rightarrow 0$ \\
        $L_{tot} \rightarrow \infty$
    \end{tabular} \\
    \hline
    
    \begin{tabular}{ c }
        $c_{fC} \rightarrow 0$ \\
        $\frac{G_{tot}\cdot c_{fGA}\cdot SA\cdot c_{tlA}\cdot \tau_{Am}}{c_{bGA}\cdot c_{fAL}} \rightarrow \infty$ \\
        $\nu \rightarrow \infty$
    \end{tabular} \\
    \hline
    
    $\frac{G_{tot}\cdot c_{fGA}\cdot SA\cdot c_{tlA}\cdot \tau_{Am}}{c_{bGA}\cdot c_{fAL}\cdot \nu} \rightarrow 0$ \\
    \hline
\end{tabular}
\caption{Sequence of limits represented by the orange arrow in Fig. 3.}
\label{tab:orangelimits}
\end{table}

\begin{table}[H]
\centering
\begin{tabular}{ |c| } 
    \hline
    Black \\
    \hline
    $c_{bC} \rightarrow 0$ \\
    \hline
    
    \begin{tabular}{ c }
        $c_{tlA} \rightarrow \infty$ \\
        $SA \rightarrow 0$
    \end{tabular} \\
    \hline
    
    \begin{tabular}{ c }
        $c_{tlA}\cdot SA \rightarrow 0$ \\
        $\tau_{Am} \rightarrow \infty$
    \end{tabular} \\
    \hline
    
    $c_{bAL} \rightarrow 0$ \\
    \hline
    
    \begin{tabular}{ c }
        $c_{fGA} \rightarrow 0$ \\
        $G_{tot} \rightarrow \infty$
    \end{tabular} \\
    \hline
    
    \begin{tabular}{ c }
        $\frac{c_{fGA}}{G_{tot}} \rightarrow \infty$ \\
        $c_{fAL} \rightarrow \infty$
    \end{tabular} \\
    \hline
    
    $\alpha \rightarrow \infty$ \\
    \hline
    
    \begin{tabular}{ c }
        $SA\cdot c_{tlA}\cdot \tau_{Am} \rightarrow \infty$ \\
        $\frac{G_{tot}\cdot c_{fAL}}{c_{fGA}} \rightarrow \infty$ \\
        $L_{tot} \rightarrow \infty$
    \end{tabular} \\
    \hline
    
    \begin{tabular}{ c }
        $c_{bGA} \rightarrow \infty$ \\
        $\frac{G_{tot}\cdot c_{fAL}}{c_{fGA}\cdot SA\cdot c_{tlA}\cdot \tau_{Am}} \rightarrow 0$
    \end{tabular} \\
    \hline
    
    \begin{tabular}{ c }
        $c_{fC} \rightarrow 0$ \\
        $\frac{G_{tot}\cdot c_{fAL}\cdot c_{bGA}}{c_{fGA}\cdot SA\cdot c_{tlA}\cdot \tau_{Am}} \rightarrow 0$ \\
        $\nu \rightarrow \infty$
    \end{tabular} \\
    \hline
    
    $\frac{\nu\cdot G_{tot}\cdot c_{fAL}\cdot c_{bGA}}{c_{fGA}\cdot SA\cdot c_{tlA}\cdot \tau_{Am}} \rightarrow \infty$ \\
    \hline
\end{tabular}
\caption{Sequence of limits represented by the black arrow in Fig. 3.}
\label{tab:blacklimits}
\end{table}

\begin{table}[H]
\centering
\begin{tabular}{ |c|c| } 
    \hline
    Original & Reparameterized \\
    \hline
    \begin{tabular}{ c }
        $c_{fC}$ \\  
        $c_{bC}$ \\
        $\alpha$ \\
        $c_{fGA}$ \\
        $c_{bGA}$ \\
        $S$ \\
        $c_{fAL}$ \\
        $c_{bAL}$ \\
        $c_{tsA}$ \\
        $c_{tlA}$ \\
        $\tau_{Am}$ \\
        $\nu$ \\
        $SA$ \\
        $G_{tot}$ \\
        $L_{tot}$
    \end{tabular} &
    \begin{tabular}{ c }
        $c_{fC}$ \\
        $c_{bC}$ \\
        $1/\alpha$ \\
        $1/c_{fGA}$ \\
        $c_{bGA}/c_{fGA}$ \\
        $S$ \\
        $c_{fAL}$ \\
        $c_{bAL}$ \\
        $\left[c_{tlA}\cdot c_{tsA}\right]$ \\
        $1/c_{tlA}$ \\
        $\tau_{Am}$ \\
        $\nu$ \\
        $\left[c_{tlA}\cdot SA\right]$ \\
        $G_{tot}$ \\
        $L_{tot}$ \\
    \end{tabular} \\
    \hline
\end{tabular}
\caption{Left column: Bare parameters. Right column: Reparameterized parameters used to generate the limits in the right hand column of Table~\ref{tab:bluelimits}.}
\label{tab:rep_limits}
\end{table}

\subsection{Supremum algorithm -- Application to Wnt signaling}
The supremum algorithm can be summarized by the following steps:
\begin{enumerate}
    \item Define the hypothesis space by selecting a complex, multiparameter model to describe the desired behaviors.
    \item Find models that minimally describe each behavior by removing parameters that can be set to 0 (or infinity) while still approximating the behavior.
    \item Using the diamond property, and possibly reparameterizing the model, find the parameter removals common to both reduced models.
    \item Apply those common parameter removals to the original, full model to obtain the supremal model.
\end{enumerate}
Here we detail this process for the Wnt signaling case from the main text.

The first step is to define the hypothesis space, i.e., the functional form of the model.
In this case, we seek a functional form containing all the mechanisms required to describe both the accumulation and oscillation behaviors.
We constructed a mass action model with a functional form based on that described in \cite{jensen2010}, with an additional controllable activation of Axin2.
This 15 parameter model (C15, Eq.~\ref{eq:C15}) contains all the mechanisms needed to describe both behaviors of interest.
This full model contains more parameters than are needed to describe all the behaviors of interest, as is often the case when defining an initial hypothesis space.

The second step in the supremum algorithm is to find the minimal (containing the fewest parameters possible) models that describes each of the behaviors of interest.
To do this we use MBAM to find a sequence of parameter limits that can be applied to the function form of the model. 
The sequence of limits used to arrive at the minimal accumulation model (A3, Eq.~\ref{eq:A3}) are shown in Table~\ref{tab:blacklimits}.
Likewise, for the minimal oscillation model (O9, Eq.~\ref{eq:O9}), the sequence of limits are shown in Table~\ref{tab:bluelimits}.

Having found minimal models, the third step is to identify the common parameter limits, i.e., those that occur in the sequence of limits of each of the minimal models.
This can be a complicated step, and care must be taken to reparameterize in a way that reveals common limits (consider the example using the diamond property in the main text).
In this case, all of the limits to derive the oscillation model occur when constructing the accumulation model, except for the green limit shown in Table~\ref{tab:redgreenlimits}, $SA\rightarrow 0$.
All of the remaining, blue limits (Table~\ref{tab:bluelimits}) occur in both set of reduction limits and are thus the ``common limits" referred to in this step.

Finally, the supremal model is constructed by applying these limits to the full model, defined in step one.
In this case, applying the blue limits from Table~\ref{tab:bluelimits} to the full model, C15 in Eq.~\ref{eq:C15}, results in model C10 in Eq.~\ref{eq:C10}, the supremal model.
This model now contains the minimal number of parameters needed to describe both behaviors of interest: the accumulation and oscillation behaviors.

The supremal model contains ten parameters, one more than is needed to minimally describe the oscillation behavior, and seven more than are needed to minimally describe the accumulation behavior.
As a result, when being fit to either behavior, the supremal model will contain more degrees of freedom than are needed, i.e., the model will contain sloppy parameters.
The supremal model can match the accuracy of models A9 and O9 individually, by using the parameter values of each model, respectively, with the additional parameters ($c_{tSA}$ for accumulation and $SA$ for oscillation) being set to zero.

\begin{figure}[h!]
    \centering
    \includegraphics[width=\textwidth]{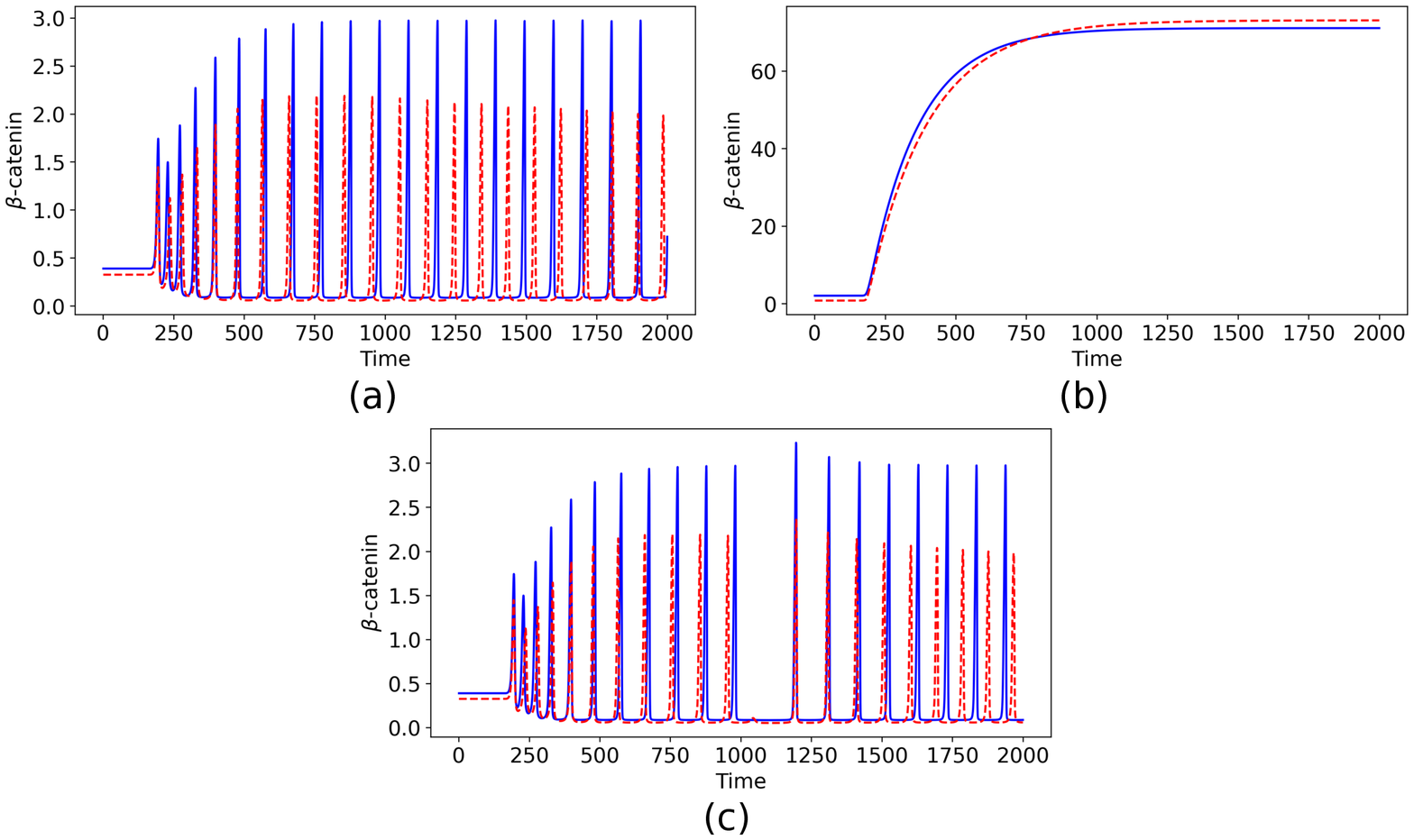}
    \caption{
    Simultaneous fit of the accumulation and oscillation behaviors with the C10, supremal model in Eq.~\ref{eq:C10}.
    This is done as shown in Eq.~\ref{eq:simfit} where the parameter $c_{tSA}$ is set to zero for the accumulation behavior, and $SA$ is set to zero for the oscillation behavior.
    (a) and (b) show fits to the accumulation and oscillation data, respectively.
    Using the fitted parameters from the simultaneous fit, with all parameters non-zero, the supremal model qualitatively reproduces the novel, interrupted behavior in (c).
    In each plot, the data are shown as a blue solid line and the model predictions as a red dashed line.
    }
    \label{fig:simfit}
\end{figure}

The parameters of the supremal model can be found by fitting both behaviors simultaneously.
This is done by solving the minimization problem
\begin{equation}
    \min\limits_{\vec{\theta}, SA, c_{tSA}} \left\{\left(f(\vec{\theta}, SA, c_{tSA}=0) - y_A\right)^2 + \left(f(\vec{\theta}, SA=0, c_{tSA}) - y_O\right)^2\right\},
    \label{eq:simfit}
\end{equation}
where $\vec{\theta}$ represents the parameters common to models A9 and O9 (all except $c_{tSA}$ and $SA$), $f$ represents the supremal model, and $y_A$ and $y_O$ represent the accumulation and oscillation data, respectively.
This simultaneous fit is demonstrated for the Wnt model in Fig.~\ref{fig:simfit}a-b.
This set of parameters also provides a qualitative fit to the novel, interrupted behavior, as shown in Fig.~\ref{fig:simfit}c.

In this example, all of the parameters in the supremal model are constrained by fitting the data oscillatory and accumulation regimes in Eq.~\eqref{eq:simfit}.  
In general, it is possible that the supremum algorithm may reintroduce parameters that are not constrained by either data set.
This will happen if the reduced models involved contradicting limits (e.g. $\theta \rightarrow 0$ in one reduction and $\theta \rightarrow \infty$ in another).
The parameter will be retained in the supremum, but it will not be identifiable by either data set.  
In this case, the supremal model will have one unidentifiable when fit to data from the two regimes.  
When this occurs, the supremum construction identifies which information is missing in the extant data that would be required to constrain future predictions. 
At this point a modeler could use optimal experimental design to look for additional data to constrain the remaining unidentifiable parameter.

\section{Additional supremum-building algorithm}
The main text presents an algorithm for finding the supremal model: given the parameter limits to arrive at two reduced models, find and apply the common limits to the original full model.
We now present a second algorithm that is less straightforward, but equally valid.
Given two reduced models, the supremum can be created by taking the union of their relevant mechanisms.
It may not be clear how to do this in all cases. However, in the Wnt example this means simply to take the union of the terms in each equations.
For example, taking the union of Eqs.~\ref{eq:A9} and \ref{eq:O9} will give Eq.~\ref{eq:C10}.
Since the only difference in Eqs.~\ref{eq:A9} and \ref{eq:O9} is in the dynamic variable $\widetilde{y}_6$ we will demonstrate this.

The dynamic equations for $\widetilde{y}_6$ for models A9 and O9 are, respectively,
\begin{align}
    \dot{\widetilde{y}_6} &= -1/\tau_{Am} \widetilde{y}_6 + \left[SA\cdot c_{tlA}\right] \\
    \dot{\widetilde{y}_6} &= -1/\tau_{Am} \widetilde{y}_6 + \left[c_{tsA}\cdot c_{tlA}\right] y_3^2.
\end{align}
Taking the union of terms in these equations gives the dynamic equation for C10, the supremal model
\begin{align}
    \dot{\widetilde{y}_6} &= -1/\tau_{Am} \widetilde{y}_6 + \left[c_{tsA}\cdot c_{tlA}\right] y_3^2 + \left[SA\cdot c_{tlA}\right] f_U(t).
\end{align}

This algorithm can be used when there is a clear connection between the dynamic variables of the reduced models, but it may not be as clear how to apply it otherwise.
Additionally, there is no need to know the full model to use this algorithm.

\bibliography{refs}